\pdfoutput=1
\newcommand{\pd}[2]{\frac{\partial #1}{\partial #2}}

\newcommand{\pdl}[2]{\partial #1 / \partial #2}

\newcommand{\Sec}[1]{Section\ \ref{#1}}

\newcommand{\Eq}[1]{Eq.\ (\ref{#1})}
\newcommand{\Eqs}[2]{Eqs.\ (\ref{#1}) and (\ref{#2})}
\newcommand{\Fig}[1]{Fig.~\ref{#1}}
\newcommand{\Figs}[2]{Figs.\ \ref{#1} -- \ref{#2}}

\newcommand{\lr}{\left(}
\newcommand{\rr}{\right)}

\newcommand{\ls}{\left[}
\newcommand{\rs}{\right]}
\newcommand{\lc}{\left\{}
\newcommand{\rc}{\right\}}

\newcommand{\avG}[1]{\langle #1 \rangle_\Gamma}

\newcommand{\be}{\begin{equation}}
\newcommand{\blnm}{\begin{linenomath*}}
\newcommand{\ee}{\end{equation}}
\newcommand{\elnm}{\end{linenomath*}}

\newcommand{\bspl}{\begin{split}}
\newcommand{\espl}{\end{split}}
\newcommand{\bea}{\begin{eqnarray}}
\newcommand{\eea}{\end{eqnarray}}
\newcommand{\bagn}{\begin{align}}
\newcommand{\eagn}{\end{align}}
\newcommand{\bs}{\begin{split}}
\newcommand{\es}{\end{split}}
\newcommand{\bc}{\begin{center}}
\newcommand{\ec}{\end{center}}
\newcommand{\bp}{\begin{picture}(0,0)}
\newcommand{\ep}{\end{picture}}
\newcommand{\bfl}{\begin{flushleft}}
\newcommand{\efl}{\end{flushleft}}

\newcommand{\bx}{\mathbf{x}}
\newcommand{\bxG}{\mathbf{x}_\Gamma}

\newcommand{\bu}{\mathbf{u}}

\newcommand{\bw}{\mathbf{w}}

\newcommand{\psixt}{\psi  \left( \bx,t \right)}

\newcommand{\Ip}{\mathcal{I \lr \psi \rr}}

\newcommand{\bnG}{\mathbf{n}_\Gamma}
\newcommand{\bng}{\mathbf{n}_\gamma}

\newcommand{\bg}{\mathbf{g}}
\newcommand{\eph}{\epsilon_h}
\newcommand{\ephb}{\epsilon_{h,b}}
\newcommand{\ephm}{\epsilon_{h,max}}
\newcommand{\ephxt}{\epsilon_h \lr \bx,t \rr}

\newcommand{\ephppxt}{\epsilon_h^{\psi'} \! \lr \bx,t \rr}
\newcommand{\ephtpxt}{\epsilon_h^{t' \psi} \! \lr \bx,t \rr}

\newcommand{\alpp}{\alpha \lr  \psi \rr}
\newcommand{\alppd}{\alpha_d \! \lr  \psi \rr}

\newcommand{\deltaa}{\tilde{\delta} \lr  \alpha \rr}

\newcommand{\psia}{\psi  \left( \alpha \right)}
\newcommand{\psix}{\psi  \left( \bx,t \right)}

\newcommand{\colw}{\color{white}}

\RequirePackage{fix-cm}
\documentclass[twocolumn]{svjour3}          
\smartqed  
\usepackage{lipsum,mathtools,cuted}
\usepackage{graphicx}
\usepackage{bbding}
\usepackage{upgreek}
\usepackage{url}
\usepackage{color,xcolor}
\usepackage{amsmath}
\usepackage{bm}
\usepackage[colorlinks,
            linkcolor=blue,
            anchorcolor=blue,
            citecolor=blue,
            urlcolor=blue,
            ]{hyperref}
\usepackage{mathptmx}      
\usepackage{titletoc}
\usepackage{amssymb}
\usepackage{flushend}

\usepackage{multicol}

\emergencystretch=\maxdimen
\titlecontents{section}[0pt]{\addvspace{2pt}\filright}              {\contentspush{\thecontentslabel\ }}              {}{\titlerule*[8pt]{.}\contentspage}

\hypersetup{
    bookmarks=false,         
    colorlinks=true,       
    linkcolor=blue,          
    linkbordercolor={1 1 1}
}

\journalname{Acta Mech. Sin.}
\begin{document}
\rmfamily
\title{Topological changes due to non-equilibrium effects\newline by means of the statistical model of two-phase flow
\thanks{This work was supported by the grant
	of National Science Center, Poland
	(Narodowe Centrum Nauki, Polska)
	in the project 
	\emph{``Statistical modeling of turbulent two-fluid flows with interfaces''}
	(ref. no. 2016/21/B/ST8/01010, ID:334165).}
}
\subtitle{{\it Faculty of Power and Mechanical Engineering, Nowowiejska 24, Warsaw 00-665, Poland }}

\titlerunning{Topological changes due to non-equilibrium effects by means of the statistical model of two-phase flow}        

\author{          Tomasz Wac{\l}awczyk 
       }


\institute{{\Envelope} corAuthor 1 \at
            \email{tomasz.waclawczyk@pw.edu.pl} \at \at
		    Faculty of Power and Mechanical Engineering, Nowowiejska 24, Warsaw 00-665, Poland \at \at
           }
\date{\copyright {\it Acta Mechanica Sinica}, The Chinese Society of Theoretical and Applied Mechanics (CSTAM) 2020
}


\maketitle


\begin{abstract}
  This 
  paper presents
  the first results 
  of the two-phase
  flow obtained
  using recently
  introduced 
  physical, mathematical and numerical 
  model of the intermittency region
  between two-phases 
  (Wac{\l}awczyk 2017, 2021).
  The 
  statistical
  interpretation
  of the intermittency
  region evolution
  equations allows
  to account for
  the non-equilibrium
  effects in
  the domain separating
  two phases.
  The source of
  non-equilibrium      
  are spatial 
  variations in
  the ratio of work
  done by volume 
  and interfacial
  forces 
  governing
  its width.
  As the statistical description
  of the two-phase flow
  differs from the
  deterministic
  two-phase flow models
  known in the literature,
  in the present work
  we focus discussion
  of the results 
  on these differences.
%
 %
  To this goal,
  the rising two dimensional 
  gas bubble 
  is studied;
  differences
  between equilibrium
  and non-equilibrium
  solutions are 
  investigated.
  It is
  argued 
  the statistical description
  of the intermittency region
  has potential to account for 
  physical phenomena not
  considered  previously
  in the
  computer
  simulations
  of two-phase
  flow.
%
\keywords{multiphase flow 
	     \and phase field method 
	     \and statistical interface model 
	     \and surface density model 
	     \and non-equilibrium phenomena in two-phase flow }
\end{abstract}
\vspace{1 cm}

%
\section{Introduction}
\label{Sec1}
%

The main difference
between single and
multi-phase flow is
in the topological
changes of the 
boundaries separating
coexisting phases 
and/or flow 
domain(s).
The wide
spectrum of
physical
phenomena
at these 
boundaries 
governs
the multi-phase flow
and have 
to be 
accounted for
in all
multi-phase 
flow
regimes.
The numerical solution 
of the flow problems
with moving boundary
conditions is source 
of complexity of the
numerical algorithms
and has led to the
development of the
abundance of 
the computational
techniques
\cite{anderson1998,osher03,prosperetti06,trygg11}.

In fluid mechanics,
the most popular
approach
splits
the control
volume  containing
neighboring phases
with the sharp
interface 
smooth enough
to compute
mean Gaussian
curvature
needed to determine
capillary terms
\cite{osher03,trygg11}.
While the division
of the given control
volume
is not arbitrary
(as one would expect
in the continuum mechanical model),
the sharp interface
model guarantees
the highest
possible numerical
resolution
of the interface
(at worst in the single
control volume and its
nearest neighbors).

In the phase field methods,
the control volume
is divided
by the diffusive interface
\cite{anderson1998}.
The main advantage
of the phase field 
methods is in their
physical interpretation
based on theory of
the Ginzburg-Landau 
free energy functional \cite{kim12}.
However,
the order
parameter 
in the phase field
model does not
have clear 
physical interpretation.
This is caused by
the fact 
Allen-Cahn \cite{allen1979} 
and Cahn-Hilliard \cite{cahn1958}
equations governing
the phase field 
does not
have a mathematical
form of 
the classical transport
equation.
%
To overcome 
these difficulties
and preserve 
conservation
of the order parameter
the Lagrange multiplayer 
techniques has been
adopted
\cite{kim12,kim14}.

In the fluid 
mechanics 
community
it is assumed,
that the numerical
sharp and 
diffusive
interface 
models
follow the thermodynamic 
dividing plane model of Gibbs \cite{gibbs1874}
and the smooth transition region
model of van der Waals
\cite{waals1979}.
We  note,
that if this  
indeed is the case,
they 
inherit 
the known 
shortcomings
of these models, too.
For instance,
as the position of the
Gibbs dividing plane $\bxG\,[m]$ is
arbitrary \cite{lang15,faust2018}
the (signed) distance 
measured from 
this position 
is not defined
as well.
In the phase 
field models
it is often
assumed the radius of 
the curvature $R\,[m]$ 
has to be much
larger than the interface 
thickness $\eph\,[m]$.
Otherwise a
numerical robustness
 and
physical interpretation
of the  phase 
field model(s) 
order
parameter are
impaired
\cite{kim12,lang15}.
As phase transitions 
(e.g. boiling)
typically 
start at the molecular
level from
the mixture of
fluid and its vapor
(no bubble exists at
this stage, yet)
the assumption
about $\eph/R\!\ll\!1$
ratio
seems to
restrict
the range 
of the phase 
field model(s)
application.

Whilst 
the mathematical and
following physical
imperfections
of the aforementioned
model equations
can be compensated 
by a careful choice 
of the  sophisticated numerical
techniques
(e.g. higher-order advection schemes \cite{liu1994},
      the height function method \cite{cummins05}, 
	  adaptive mesh refinement \cite{popinet09},
      Lagrange multiplayer \cite{kim14},
      etc.)
leading, eventually, to plethora
of successful numerical predictions;
the most striking flaw of 
the deterministic sharp/diffusive
interface models  seems 
to be their detachment from 
the experimental reality
\cite{vrij1973,aarts2004}.

Moreover, 
one could also ask:
how likely is it ,
that solution of
the  five \cite{cahn1958,allen1979,hirt81,osher1988,olsson05},
different (sets of) 
apparently unrelated 
partial differential
equations,
describing 
the same physical phenomenon(a),
using different
boundary conditions,
leads to the same 
(or very similar) predictions?

The answers 
to some of these  questions
are given in \cite{twacl17}.
Unlike described by the sharp
and diffusive interface models
the mesoscopic interface $\Gamma$
between two phases
is rough and/or turbulent.
It is disturbed
by thermal fluctuations
in the domain where
particles 
of coexisting
phases violently
exchange energy
dependent on
the temperature
and cohesive forces of 
the given system.
For this reason, 
the width of 
this domain is
estimated to be
$\eph \!\sim\! \sqrt{k_B T/\sigma} \! > \! 0$
where
$k_B\,[J/K]$ is 
the Boltzman constant,
$T\,[K]$ is
the absolute 
temperature and
$\sigma\,[J/m^2]$ is 
the surface tension
coefficient
\cite{vrij1973,aarts2004}.
The discrepancy between
true nature of the interface
and its smooth, deterministic models
has
led the present author to
development of the description
that is alternative for the
sharp and diffusive interface
models and yet it shows
they can be considered as
complementary
components 
of the same
statistical description 
\cite{twacl15,twacl17,twacl21}.

In the present work,
recently
derived \cite{twacl15}
equation describing
evolution of 
the intermittency region
is coupled with
the Navier-Stokes 
incompressible
flow
solver  
and used to
simulate 
flow of
the rising, 
two-dimensional
gas bubble.
In addition,
we use
non-equilibrium
solution
of the modified
Ginzburg-Landau
functional \cite{twacl17,twacl21}
derived by 
the present author
to introduce
the new mechanism
of the 
topological changes
in two-phase flow.
To this goal,
the two cases are considered:
first, when characteristic
length scale field $\ephxt$
 governing
the local intermittency region
width is constant,
and second, 
when $\ephxt$
is given by 
the prescribed
analytical
function.

This
paper is organized
as follows.
In the next section,
derivation of the
intermittency region
evolution equation
and its
relation with
the modified
Ginzburg-Landau 
free energy 
functional
is recalled  
\cite{twacl17,twacl21}.
Next, we 
present the
numerical
method and
set-up of
the two-dimensional
numerical
experiment.
Finally,
obtained results
are discussed
in context of
difference between
equilibrium and
non-equilibrium 
effects in
two-phase flows
and differences
between the predictions
of the deterministic
and the statistical
intermittency
region models.

\section{Statistical model of the intermittency region}
\label{Sec2}

In this section we briefly recall
arguments already presented in  
works \cite{twacl15,twacl17,twacl21}.
Therein the derivation,
physical interpretation
and convergence
of numerical solution 
of 
the  intermittency region
evolution equation
is investigated. 
The new contribution in
these work is description
of the intermitency region
in terms of probability of finding
the mesoscopic sharp interface $\Gamma$.
Moreover,
establishing its relation to
the minimization procedure
of the modified Ginzburg-Landau 
free energy functional.

%

\subsection{Derivation of the cumulative distribution function transport equation}
\label{ssec21}

We  
assume,
the interface
between two-phases is
domain named 
the microscopic intermittency region.
Therein, the
position of the sharp interface $\Gamma$
(herein defined on the molecular level)
can be found with 
the non-zero probability 
described by the cumulative
distribution function $\alpha\,[-]$.
It is important
to note, 
the intermittency 
region paradigm
was first introduced and 
developed for modeling of
turbulence interface interactions
\cite{brocchini01a,mwaclawczyk11,waclawczyk2015}.
In \cite{twacl17} 
we have drew  analogy
between the interface
interacting with turbulence
and sharp interface $\Gamma$
agitated by
thermal fluctuations.
In both cases, 
these phenomena
can be described
as stochastic
processes.
This allowed us 
to use the conditional
averaging procedure \cite{pope88}
and eddy viscosity
model \cite{waclawczyk2015}
to close unknown
correlations
as in
turbulence-interface
interaction model.
As a result,
equation
for evolution
of the probability
of finding the
mesoscopic sharp interface
$\Gamma$: $\alpha\,[-]$, 
 is derived
\be
\pd{\alpha}{t} \!+\! \bw \nabla \alpha 
\!=\! \nabla \cdot \lr D \nabla \alpha \rr 
-  |\avG{\bnG}| \nabla{\lr D \Sigma \rr}  \!\cdot\! \bng
\label{eq1}
\ee
where $\bw\,[m/s]$ 
is  velocity of
the regularized
interface $\gamma$
defined by the expected
position of $\Gamma$: $\alpha \lr \psi\!=\!0 \rr \!=\!1/2$ , 
$\bng$ is vector
normal to $\gamma$,
$\bnG$ is vector
normal to $\Gamma$,
$\avG{\cdot}$ is
the conditional average operator
and 
$D\!=\!C \lr \bx,t \rr \ephxt\,[m^2/s]$
is the diffusivity
coefficient.
$C \lr \bx,t \rr$,  
$\ephxt$
are velocity and
length scales
characterizing
the intermittency
region, respectively.
The first RHS term in \Eq{eq1}
is responsible for spreading
of the intermittency
region width, 
the second
one has been shown to be
the counter gradient
diffusion 
responsible 
for its contraction
\cite{mwaclawczyk11,waclawczyk2015}.

As \Eq{eq1} contains
an unclosed RHS term with 
the unknown
surface density 
$\Sigma\,[m^2/m^3]$,
we have used results
in \cite{cahn1958,olsson05,chiu11}
for its conservative closure
and its physical interpretation \cite{twacl21}.
Hence, $\alpha\,[-]$ evolution equation reads
\be
\pd{\alpha}{t} \!+\! \nabla \! \cdot \! \lr \bw \alpha \rr \!=\! \nabla \cdot \ls D |\nabla \alpha|\bng \!-\! C \alpha \lr 1-\alpha \rr \bng \rs.
\label{eq2}
\ee
The coefficients $C\,[m/s],\, D\,[m^2/s]$
in \Eq{eq2} specify
the characteristic
length $\eph \!\sim\! D/C\,[m]$ 
and time $\tau_h \! \sim \! D/C^2\,[s]$
scales 
governing its
solution
and in general
can be 
functions of
space and time. 
Thus,
in works \cite{kmtw14,twaclawczyketal14,waclawczyk2015}
it was shown \Eq{eq2} can
be used to predict 
the intermittency region
evolution due to interaction
of turbulent eddies with
the macroscopic
interface $\gamma$ 
(the same interface as 
in the standard volume of fluid (VOF) 
and level-set (SLS) 
numerical models).

The steady
state
solution 
of \Eq{eq2}
with $\eph\!=\! const.$
and
$\bw\!=\!\bu \!=\!0$
is given by
the regularized
Heaviside function 
\be
\alpp = \frac{1}{1+\exp{\lr -\psix/\eph \rr}}=
\frac{1}{2} \ls 1+\tanh{\lr \frac{\psix}{2\eph} \rr} \rs
\label{eq3}
\ee
and its inverse
function that is
the signed distance 
from the expected position of 
the regularized interface $\gamma$ 
defined by the level-set $\psi \lr \alpha \!=\!1/2 \rr \!=\! 0$
\be
\psi \lr \alpha \rr = \eph \ln{\ls \frac{\alpha\lr \psi \rr}
	{1 - \alpha \lr  \psi \rr} \rs}.
\label{eq4}
\ee
As noticed
by the 
present author
\cite{twacl15},
\Eqs{eq3}{eq4} 
are known 
to characterize
the cumulative 
distribution $\alpp$,
and quantile $\psia$
functions
of the logistic
distribution.
Additionally,
the gradient 
of $\alpp$ 
given by the formula
\be
\nabla \alpha = \frac{\deltaa}{\eph} \nabla \psi,
\label{eq5}
\ee
where $\deltaa/\eph \!=\! \alpha \lr 1\!-\!\alpha \rr/\eph$ is 
the probability density function
of the logistic distribution.
Computation of $\nabla \alpha\,[1/m]$
with \Eq{eq5} 
bridges
the numerical problems
with approximation
of steep functions
gradients (see $\alpha$ when $\eph \rightarrow 0$ and analysis in  \cite{twacl15}).
\Eq{eq5} can also be interpreted as
calculation of $\nabla \alpha$ 
in the $\bng$ direction 
\be
\nabla \alpha \cdot \bng \!=\! \pd{\alpha}{\psi} \!=\! \frac{\deltaa}{\eph}.
\label{eq6}
\ee
This observation was
recently used in \cite{holger2021}
to reformulate the
modified Ginzburg-Landau
free energy functional
for its solution
in the local 
coordinate system
normal in every
point to 
the interface
$\gamma$.
While \Eq{eq6} is exact,
the Laplacian 
$\nabla \!\cdot\! \lc \nabla \ls \alpp \!\cdot\! \bng \rs \rc$ 
in the normal direction 
can be only approximated
in similar way
when the interface
curvature $\kappa \!\ll\! 1$.

Curiously,
the definition 
of the chemical potential 
and the corresponding 
modified Ginzburg-Landau
functional 
in \cite{twacl17} and \cite{holger2021} 
is the same (to details 
in the coefficients and the fact
that different solution norms
are used \cite{mirja20}).
The advantage 
of description
used herein
over the classical phase 
field models
is  sound physical
interpretation
of the order
parameter $\alpha\,[-]$, its gradient $\nabla \alpha\,[1/m]$
and the order parameter inverse function $\psi\,[m]$, 
see Eqs.~(\ref{eq3},\ref{eq5},\ref{eq4})
respectively.

This makes the difference
as for instance,
one can use the physical 
meaning of the cumulative
distribution function given
by \Eq{eq3}
to pinpoint 
the position 
of the Gibbs dividing
surface $\bx_\Gamma$.
The Gibbs 
dividing surface
can be
defined 
as the expected
position of 
the sharp interface $\Gamma$
agitated by
the stochastic 
thermal fluctuations.
This position 
is given by the level-sets
$\alpha \lr \psi\!=\!0 \rr\!=\!1/2$
and describes 
the two dimensional
smooth surface that
is called
the sharp interface
in the deterministic
models of
the intermittency 
region.

In \cite{twacl15}
it was noticed that
substitution of \Eq{eq5}
into \Eq{eq2} results in
\be
\pd{\alpha}{t} + \bw \nabla \alpha =  \nabla \cdot \ls C  \deltaa \lr |\nabla\psi|-1 \rr \bng \rs,
\label{eq7}
\ee
where 
$\bng \!=\! \nabla \alpha /|\nabla \alpha|\!=\! \nabla \psi /|\nabla \psi|$.
In the present work
we separate the advection 
and re-initialization
steps in \Eq{eq7},
which leads to
\be
\pd{\alpha}{t} \!+\! \bw \nabla \alpha \!=\!
\pd{\alpha}{t} \!+\! \frac{\deltaa}{\eph} \bw \!\cdot\! \nabla \psi  \!=\! 0,
\label{eq8}
\ee 
\be
\pd{\alpha}{\tau} \!=\!  \nabla\! \cdot \! \ls C  \deltaa \lr \left |\nabla\psi \right| \!-\! 1 \rr \bng \rs,
\label{eq9}
\ee
where $\tau\,[s]$ 
is time needed to
obtain the equilibrium
solution of \Eq{eq9}.
In our works, 
this form
of \Eq{eq2}
is preferred
for numerical
solution
and theoretical
analysis.

At first,
the choice of 
the partial
differential algebraic
equation (\ref{eq4}-\ref{eq7})
for the model
of two-phase
flow in fluid
mechanics appears
to be surprising.
However,  
it is known \cite{ascher1998},
the equations of this type  
describe systems with
the large separation of 
scales (e.g. chemical kinetics in quasi steady state 
        and partial equilibrium approximations,
        molecular dynamics, optimal control etc.).
This suits our goal
of derivation of the
intermittency region model
that is closer to
experimental reality
than known  in the literature
thermodynamic 
descriptions of
the interface
between two-phases.

\subsection{Re-initialization as minimization of the modified Ginzburg-Landau functional}
\label{ssec22}

In  \cite{twacl21} it was shown,
when the 
characteristic
length $\eph\,[m]$
and velocity 
$C\,[m/s]$
scales
governing evolution 
of the intermittency
region are constant,
this region is remaining
in the equilibrium 
state according
to the modified
Ginzburg-Landau 
functional describing
interfacial energy
of the intermittency
region \cite{twacl17}.
In this case, 
the solution 
of \Eq{eq9} in the Conservative Level-Set method (CLS) \cite{olsson05}
is called 
re-initialization
of the function $\alpha$.
In this
numerical method,
the re-initialization
step is used to restore
deformed due to numerical
errors in the advection step
function $\alpha$
to its original
form given by \Eq{eq3}.

In the limit $\eph \! \rightarrow \! 0$,
\Eqs{eq5}{eq9}
can be used to  derive
the re-initialization equation
of the signed-distance function $\psi$
used in the Standard Level-Set (SLS) method
\cite{twacl15,twacl17}.
Here,
the goal of 
the re-initialization step
is similar.
The goal function 
in reconstruction 
procedure is $\psi\,[m]$ disturbed 
by the numerical
errors introduced
during 
the advection
step
\cite{osher03}.
\Eqs{eq4}{eq7} explain
that these two functions
(and two models of
the intermittency
region)
are closely related.
Moreover,
the present
author 
has proven \cite{twacl17},
the re-initialization
procedure 
in the SLS \cite{osher03}
and CLS \cite{olsson05} level-set
methods
are equivalent
to minimization
of the 
free
energy
functional
containing 
the term 
which accounts
for
the regularized 
interface $\gamma$ 
deformation.

We note  
that before
work \cite{twacl17}
the presence
of term
with $\nabla \alpha$
and the interface
curvature $\kappa$
in 
the definition
of chemical potential
was postulated \cite{folch1999}.
Extension of
the original 
free energy
Ginzburg-Landau functional
by the additional term containing 
the interface curvature
was also proposed
in \cite{jamet2008}. 
However,
as noticed 
in  \cite{mirja20}
the derivations
in \cite{folch1999,jamet2008}
and \cite{twacl15,twacl17}
were carried out
in the different
norms.
Moreover,
as described 
in \Sec{Sec1} 
the present 
work is
driven by the
attempt to paint
more  
physical
and general
picture
of the interface.
In particular
when it is
agitated by 
the thermal 
or turbulent
fluctuations,
whereas
the  works \cite{folch1999,jamet2008}
use mainly
mathematical
arguments.
%

%
\subsection{Non-equilibrium of energy ratio in intermittency region}
\label{ssec23}


In most of 
the analytical and numerical 
discussions
presented in the literature  
it is assumed,
the interface
or intermittency
region are passive
actors advected 
by the fluid
velocity.
%
In our view,
the role 
of intermittency
region in the multi-phase
flow is much
more complex.
The intermittency
region is open system
that can interplay
with neighboring
phases through exchange
of the mass and/or energy.
%
%
In particular,
in the case of phase changes
it mediates in the exchange 
of mass and energy.
For these
reasons,
the intermittency 
region between
two weakly miscible
phases
may not
be in 
the equilibrium
state as it
is typically
assumed in
the known
sharp and
diffusive
interface
models.
%
%
Next
it is demonstrated,
that local
non-equilibrium
effects can
influence 
distribution
of the material 
properties
of the phases,
and hence, 
affect
the two-phase
flow scenario.

Driven 
by this 
general picture
in the recent
work \cite{twacl21}
the non-equilibrium
solution of
\Eqs{eq7}{eq8}
was introduced.
Therein we have 
shown, that statistical
model of the intermittency region
can also be 
extended to the case
when $\ephxt\!\ne\!const.$.
In particular
we have derived 
the free energy 
functional with
$\nabla \alpha \!\cdot\! \nabla \eph$
term  showing
the variation of
volume and
interfacial
forces work ratio
should affect the
cumulative distribution
function $\alpp$
shape.
Moreover,
we have demonstrated
the stationary solution 
of \Eq{eq9}
can be used 
as approximate
solution of the
non-equilibrium
problem
when used
in a spirit of
re-initialization
as in the CLS and
SLS numerical 
methods.

In what follows we repeat
derivation of the approximate
solution already presented
in \cite{twacl21}.
For mathematical details
regarding the modified Ginzburg-Landau
free energy functional
the interested reader should
refer to \cite{twacl21}.
The equilibrium
condition
obtained 
from  
the stationary
solution 
to \Eq{eq9}
reads
\be
\nabla \alpha \!=\! |\nabla \alpha| \bng \!=\! \frac{\alpha \lr 1\!-\!\alpha \rr}{\ephxt} \bng. 
\label{eq10}
\ee
\Eq{eq10} is 
formulated
in the direction $\bng$
normal to the  regularized
interface $\gamma$,
hence,
it may
be rewritten as
\be
\pd{\alpha}{\psi} \left| \pd{\psi}{\bx} \right| \!=\! \frac{1}{\ephxt}  \alpha \lr 1 \!-\! \alpha \rr,
\label{eq11}
\ee
where it 
is assumed
$\pdl{\alpha}{\psi} \! > \! 0$
meaning
$\alpp$ 
is expected 
to be 
the cumulative
distribution
function
with infinite 
support
analogously
to \Eq{eq5}.
Next, we assume
$|\nabla \psi| \!\equiv\! 1$
in \Eqs{eq10}{eq11}.
As a result,
substitution  
of \Eq{eq10}
into \Eq{eq2} with
$D \lr \bx,t \rr\!=\!C \ephxt$
let us derive 
\Eq{eq7} with
the variable
characteristic
length scale
governing width
of the intermittency
region.
The assumption
$|\nabla \psi| \!\equiv\! 1$
means
the signed distance
function $ \psix $
spans 
the space where
surface
averaged oscillations
of the sharp interface 
$\Gamma$ 
take place.
On average,
these
oscillations occur
only in  the direction
$\bng$ normal 
to the expected
position $\psi\!=\!0$
of the regularized
interface $\gamma$.
%

Further,
it is noticed
at each point 
$\lr \bx,t \rr$ 
of the field
$\ephxt$
the signed 
distance
function
$\psixt$ is given.
Hence, 
the knowledge of
the field $\psixt$
gives
$\lr \bx,t \rr$
and thus
$\ephxt$.
Therefore, 
we introduce
$\epsilon_h^{\psi} \lr \bx,t \rr$
denoting 
$\ephxt$ 
determined 
using $\psixt$.
This 
let us
to integrate
\Eq{eq11}
in the local 
coordinate
system attached
to the 
regularized
interface $\gamma$.
As $\gamma$
is defined 
by $\psixt\!=\!0$,
$\psixt$ is
the normal coordinate
with the origin at
$\psixt\!=\!0$
of this local 
system.
At each
fixed
point 
of given
$\alpp,\,\psia,\,\ephxt$
fields
this
integration
reads
\be
\int^{1/2}_{\alpp} \frac{d\alpha'}{\alpha' \lr 1\!-\!\alpha' \rr} = \int^{0}_{\psia} \frac{d \psi'}{\ephppxt}.
\label{eq12}
\ee
The integration
(\ref{eq12})
is performed
from the
arbitrary point
located at the signed-distance
from the regularized interface
$\alpp\!-\!\psia$
to the expected
position of the
regularized
interface $\psi \lr \alpha \!=\!1/2 \rr \!=\! 0$.
One notes
the LHS integration 
in \Eq{eq12}
does not
assume
or result in
any specific form/shape
of the function $\alpp$.

To reconstruct
the equilibrium
solution when
$\ephxt \!=\! const.$
it is necessary  
to preserve the mapping
between $\alpp\!-\!\psia$,
see \Eq{eq4}.
For this reason,
it is more convenient
to reformulate 
the RHS integral
in \Eq{eq12} using
a variable
substitution 
as follows 
\be
\int^{0}_{\psia} \frac{d \psi'}{\ephppxt} 
\!=\! \psia \int^0_1 \frac{dt'}{\ephtpxt} \!=\! \psia \Ip
\label{eq13}
\ee
where $t' \! \in \!  [ 0,1 ]$ 
is the  parameter
such that $\psi' \!=\! t' \psi$
and $d \psi' \!=\! dt' \psi$,
furthermore
$\Ip$ is used to denote
the integral on the RHS
of \Eq{eq13}.

After integration
of \Eq{eq11} with \Eq{eq13}
one obtains 
\be
\psia  = \frac{1}{\Ip} ln \ls \frac{\alpp}{1-\alpp} \rs.
\label{eq14}
\ee
It is noted, \Eq{eq14}
can be the
source of
numerical
problems when $\alpha \rightarrow 0$
or $\alpha \rightarrow 1$.
In these limits,
due to the finite computer
arithmetic
computation of the
natural logarithm is
ill conditioned
(see discussion of this 
problem in \cite{twacl15}) 
and
some measures must
be taken to account
for it.
In the code,
it is chosen to solve
\Eq{eq9} only in the
region where
the numerical
solution is still 
possible assuming: 
$|\nabla \psi|\!=\!1$
and hence $\pdl{\alpha}{\tau}\!=\!0$,
outside.

At
the given,
arbitrary
point $\lr \bx,t \rr$, 
the
signed
distance $\psixt$
has the known value.
For this reason,
at the point $(\bx,t)$
the integral
$\Ip \!=\! const.$
and thus
the inverse 
relation 
is also true
\be
\alpp \!=\! \frac{1}{1+\exp{ \ls - \psia \Ip \rs }}.
\label{eq15} 
\ee
In the present work,
\Eq{eq15} is called
the approximate,
semi-analytical solution
of \Eq{eq9}.
%

When
the field
$\ephxt \!=\! const.$,
\Eq{eq12} and \Eq{eq13}
reduce to
the equilibrium
solution,
which is guaranteed
by the
definition of $\Ip$.
Thus, 
the mapping given 
by \Eq{eq14} or
the form of $\alpp$
given by \Eq{eq15}
can be employed during
numerical solution
of the system given 
by \Eqs{eq8}{eq9}
to model
how
the 
$\ephxt$
field
is affecting 
changes of
the
cumulative 
distribution
function
$0 \!<\! \alpp \!<\! 1$
profile.

In what follows,
we will couple
described above  
approximate solution
of \Eq{eq9} given by \Eq{eq15}
with the Navier-Stokes
equation solver
to predict how 
non-equilibrium
affects affect 
the flow 
of
 the two-dimensional
rising gas bubble.
\begin{multicols}{1}
	\begin{figure*}[htbp]
		\begin{minipage}{.5\textwidth}
			\centering  \includegraphics[width=0.6\textwidth]{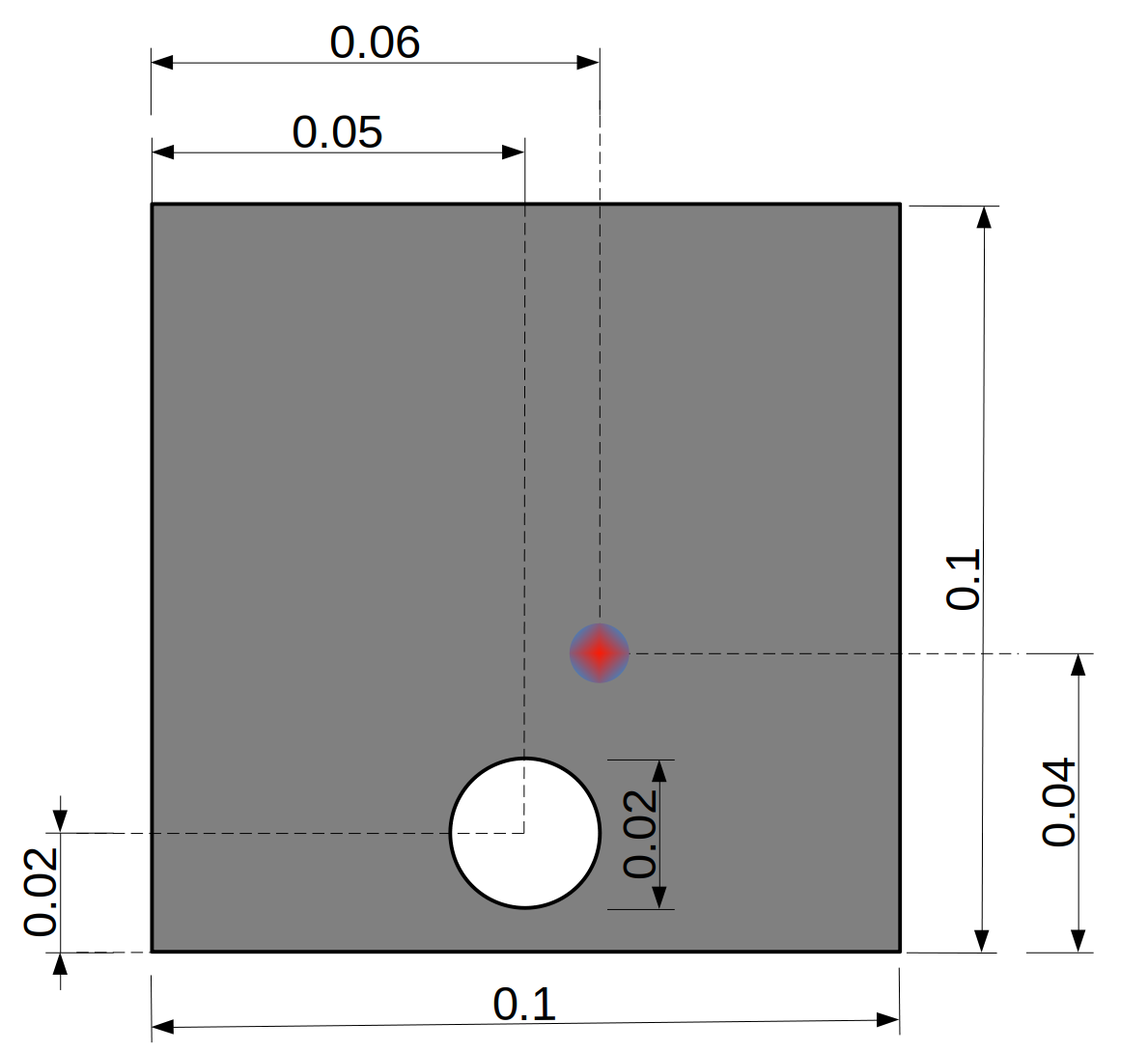}
			
			\caption{The schematic sketch of the flow case, all
				dimensions are in meters. The colored spot shows
				fixed position
				of variation in 
				the characteristic length scale field
				$\ephxt$ given by \Eq{eq19}.}
			\label{fig1} 
		\end{minipage}   
		  
	\end{figure*}
\end{multicols}
%
%
%
\section{Numerical method}
\label{Sec3}

This section describes
the numerical method used 
in the present paper
for coupling of Eqs.~(\ref{eq8}, \ref{eq9}, \ref{eq15})
with the incompressible
Navier-Stokes solver.
In addition,  
details
of incorporation
of the non-equilibrium 
effects into the 
two-phase 
flow solver
are described.

\subsection{Flow solver and one-fluid model implementation}
\label{ssec31}

In the present paper,
the second-order accurate finite
volume flow solver 
in the collocated variable
arrangement is used \cite{schaefer06,peric02}.
For detailed
description of the algorithm(s)
and presentation of the results,
see 
previous
works by the present author \cite{waclawczyk05,waclawczyk06,waclawczyk07,waclawczyk08,waclawczyk08_2,waclawczyk08_3,waclawczyk08_phd,waclawczyk2015}.
As in these
references, herein,
the incompressible
Navier-Stokes
equation 
in the conservative
form is 
solved.
The momentum
conservation 
equation is
coupled with 
the Poisson
equation to bind
 pressure
and velocity fields
in the SIMPLE 
procedure.
%
%
%
The equilibrium between
the pressure gradient and mass forces
is established using 
the same numerical 
operators for their
discretization \cite{zun07}.
In this standard solver
of incompressible flow, 
the two-phase
flow model 
using the one-fluid 
formulation \cite{trygg11}
is implemented; 
in the present
work the presence 
of the capillary
forces 
is neglected.
Material properties of 
one-fluid change according to
relations
\bea
 \rho = \rho_1 \alpha + \rho_2 \lr 1-\alpha \rr,\\
 \label{eq16}
 \mu  = \mu_1 \alpha + \mu_2 \lr 1-\alpha \rr,
 \label{eq17}
\eea
where $\rho\,[kg/m^3],\,\mu\,[Pa \!\cdot\! s]$ denote 
density and dynamic viscosity
of the mixture of
two phases; $\rho_k,\,\mu_k,\,\,k=1,2$
are   density and dynamic viscosity
of pure fluid and gas
phases, respectively.
\begin{figure*}[htbp]
	\begin{minipage}{.49\textwidth}
		\centering  \includegraphics[width=0.9\textwidth]{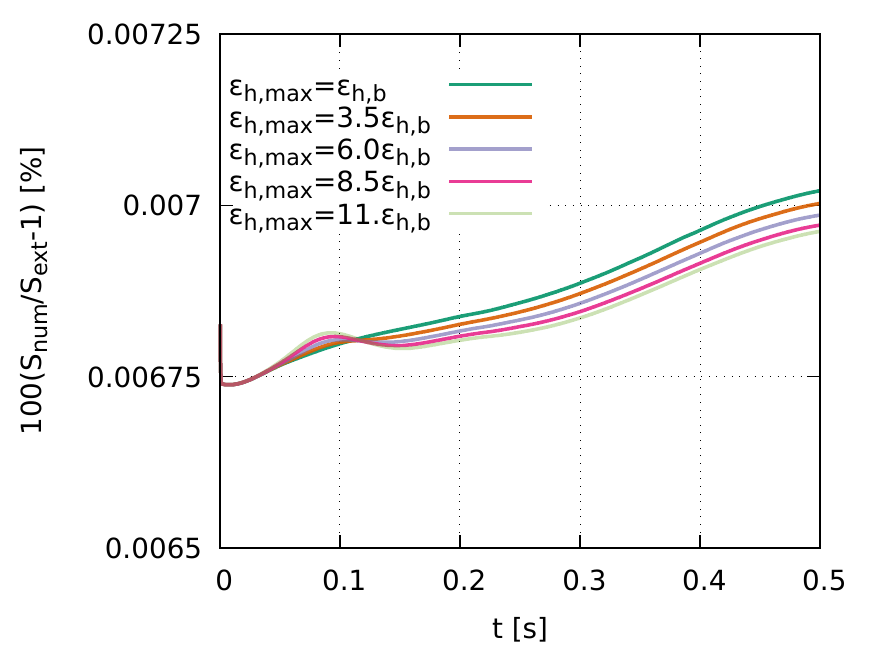}
	\end{minipage}%
	\begin{minipage}{.49\textwidth}
		\centering  \includegraphics[width=0.9\textwidth]{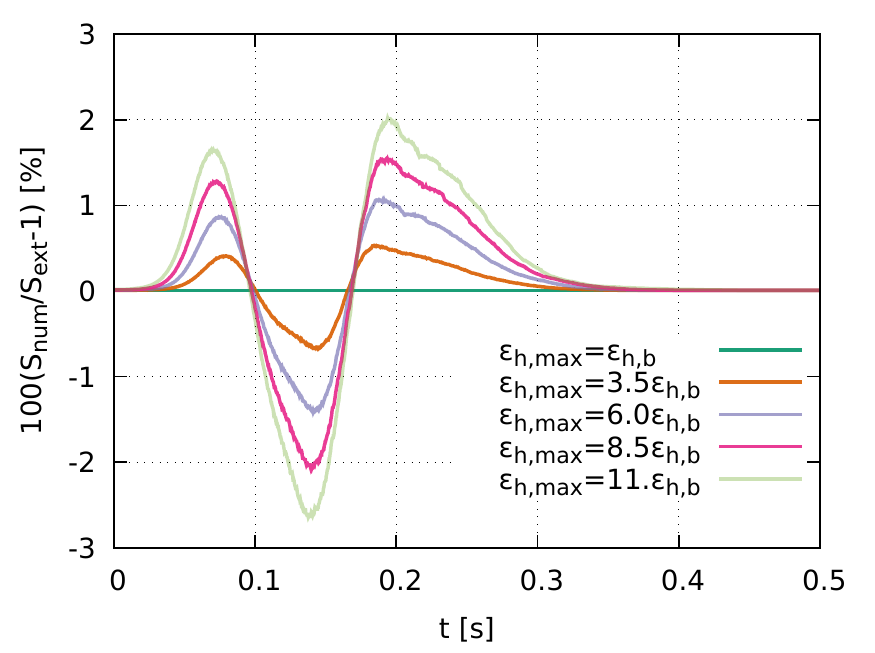}
	\end{minipage}                
	\caption{The mass conservation of gas during 
		coupled simulations with constant $\ephxt\!=\!\ephb$
		and variable  characteristic length scale field $\ephxt$.
		Total mass is computed using the c.d.f-s $\alpp$  (left)
		and $\alppd$(right).
		$\ephm$ denotes the maximum magnitude of $\ephxt$ given by \Eq{eq12}.
		}
	\label{fig5}
\end{figure*}

The advection \Eq{eq8}
is discretized in time using 
the second-order accurate implicit 
Euler method.
For discretization of
the convective term in space,
the deferred-correction 
method with 
the second-order 
accurate TVD MUSCL 
ﬂux limiter \cite{xue98}
is employed.
The same 
temporal and spatial
discretization is used
for velocity components
in the Navier-Stokes 
equation.
Re-initialization equation
(\ref{eq9}) is integrated
in time 
using the explicit third-order
accurate TVD Runge-Kutta 
method \cite{gottlieb98}.
The constrained interpolation
\cite{twacl17,twacl21} 
is used to discretize
$\deltaa\!=\!\alpha \lr 1-\alpha \rr$,
in \Eq{eq9}.
For more details
about numerical schemes
and discretization
used to solve \Eqs{eq8}{eq9},
refer to  
Appendixes in \cite{twacl15,twacl17,twacl21}.

It is noticed,
\Eq{eq7}
in the limit $\eph \rightarrow 0$ 
guarantees the statistical model
of the intermittency region 
will converge to the solution obtained
using the sharp interface model(s).
We choose
base or minimal 
width
of the intermittency
region as $\ephb\!=\!\sqrt{2}\Delta x/4$;
in  \cite{twacl15} 
it was checked this 
selection guarantees the second
order convergence of \Eq{eq9}
and its higher-order derivatives.

Furthermore it is noted,
in the  statistical model
of the intermittency region,
unlike in
some phase
field solvers
\cite{kim12},  
there
is no danger 
the material
properties 
of the fluids
will have 
nonphysical
values.
This is due
to the fact
that $\alpp$
is the cumulative
distribution
function 
(see \Eq{eq3})
and it
is incorporated
into the solution
procedure.
Therefore,
from its definition
it is always
bounded $0\!<\!\alpp\!<\!1$.
%
%
%
%
\begin{figure*}[t]
	\begin{minipage}{.16\textwidth}
		\centering\includegraphics[width=0.95\textwidth,height=1.5\textwidth,angle=0]{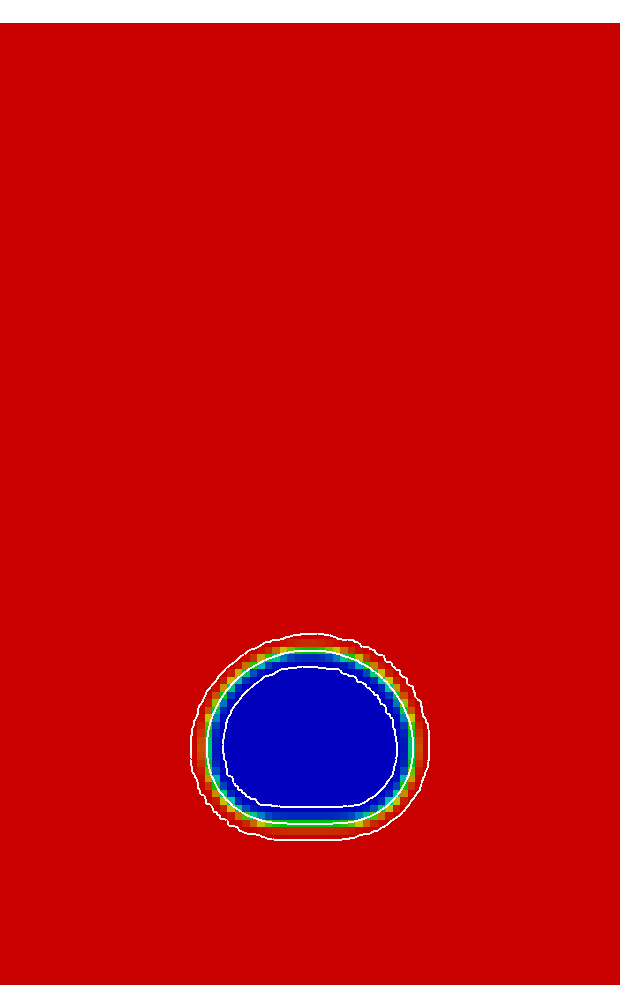}
	\end{minipage}%
	\begin{minipage}{.16\textwidth}
		\centering\includegraphics[width=0.95\textwidth,height=1.5\textwidth,angle=0]{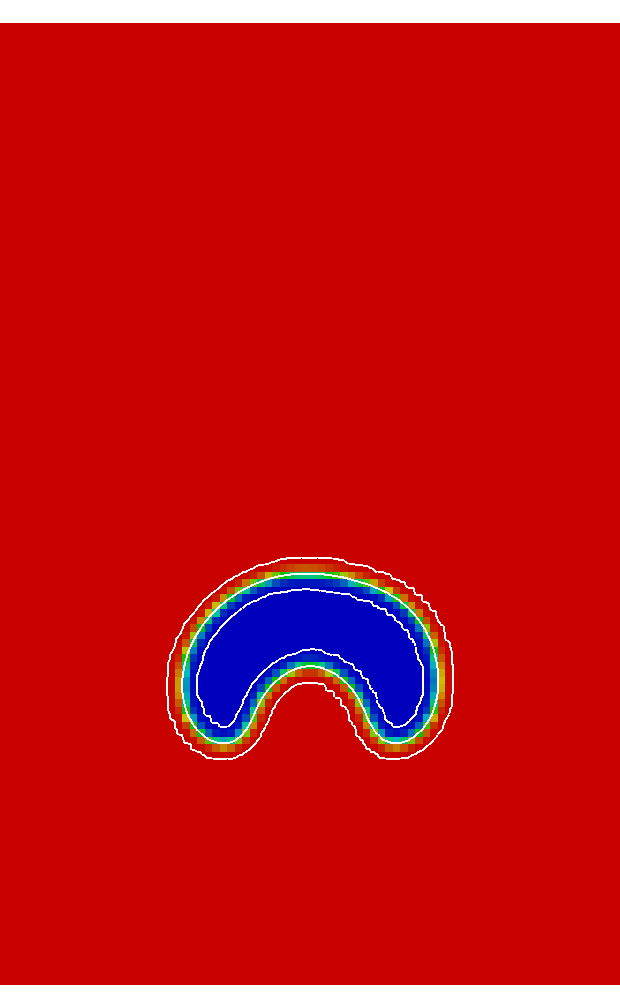}
	\end{minipage}%
	\begin{minipage}{.16\textwidth}
		\centering\includegraphics[width=0.95\textwidth,height=1.5\textwidth,angle=0]{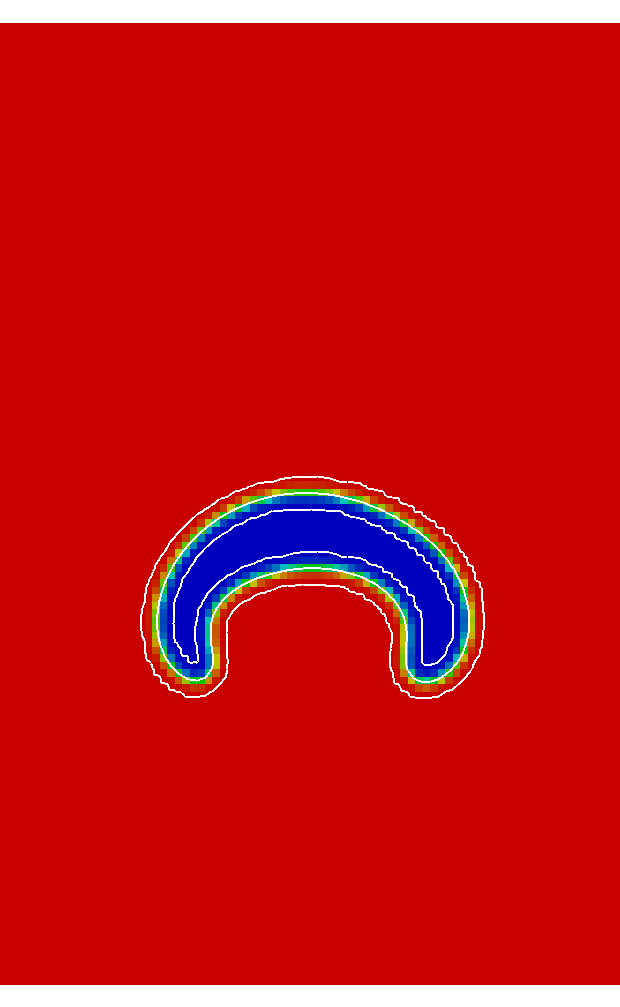}
	\end{minipage}%
	\begin{minipage}{.16\textwidth}
		\centering\includegraphics[width=0.95\textwidth,height=1.5\textwidth,angle=0]{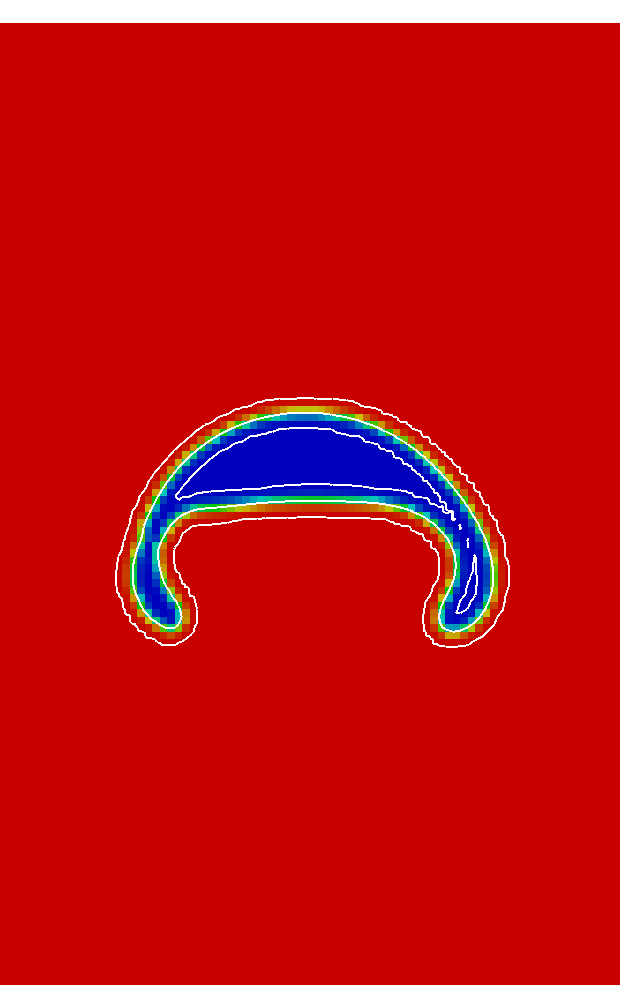}
	\end{minipage}%
	\begin{minipage}{.16\textwidth}
		\centering\includegraphics[width=0.95\textwidth,height=1.5\textwidth,angle=0]{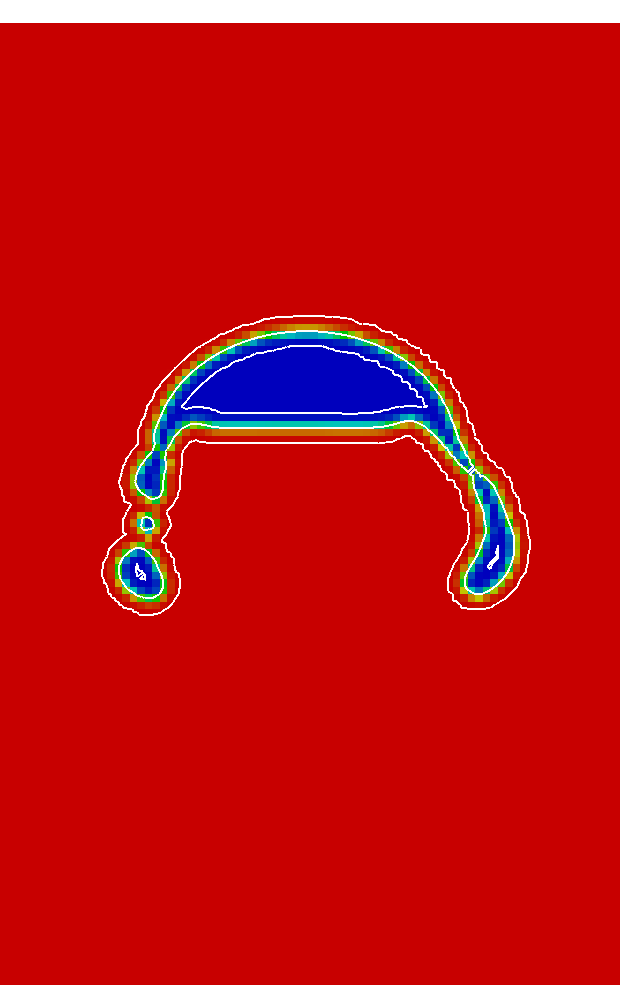}
	\end{minipage}%
	\begin{minipage}{.16\textwidth}
		\centering\includegraphics[width=0.95\textwidth,height=1.5\textwidth,angle=0]{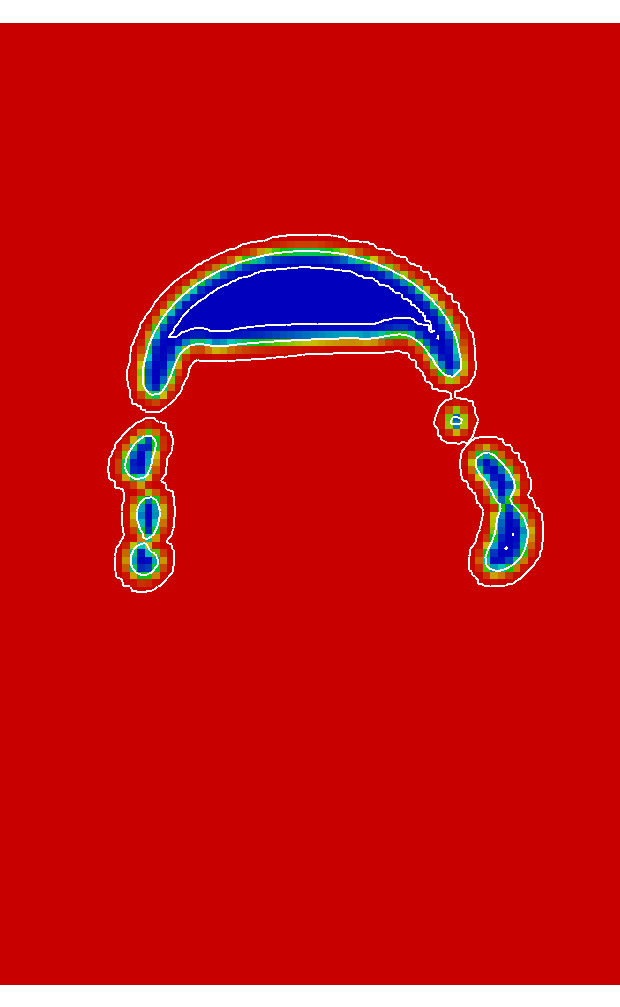}
	\end{minipage}%
	\begin{minipage}{.15\textwidth}
		\includegraphics[width=0.2\textwidth,height=1.25\textwidth,angle=0]{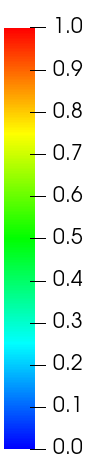}
	\end{minipage}
	\begin{minipage}{.16\textwidth}
		\centering\includegraphics[width=0.95\textwidth,height=1.5\textwidth,angle=0]{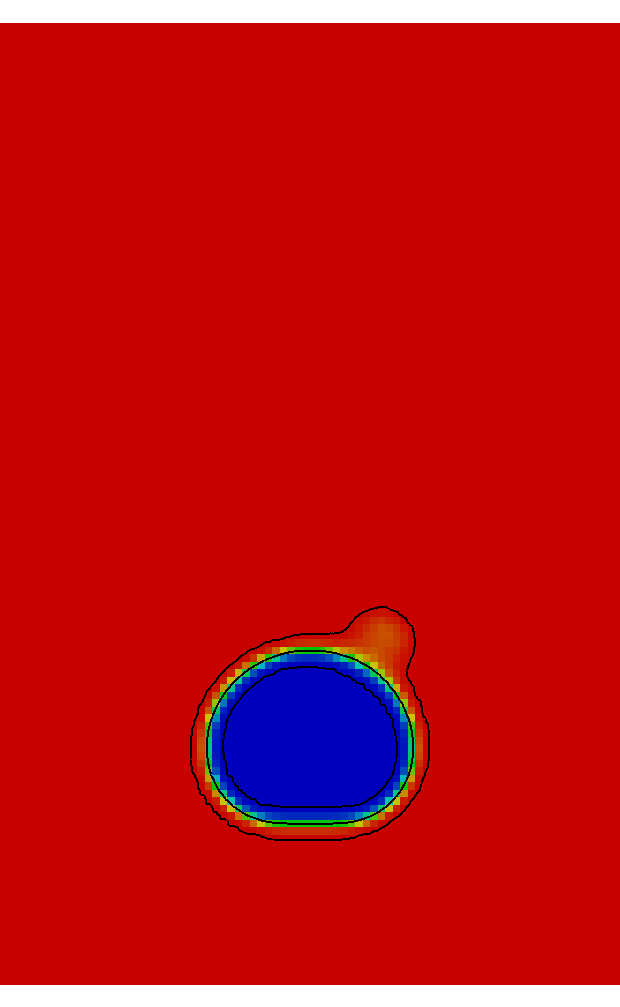}
	\end{minipage}%
	\begin{minipage}{.16\textwidth}
		\centering\includegraphics[width=0.95\textwidth,height=1.5\textwidth,angle=0]{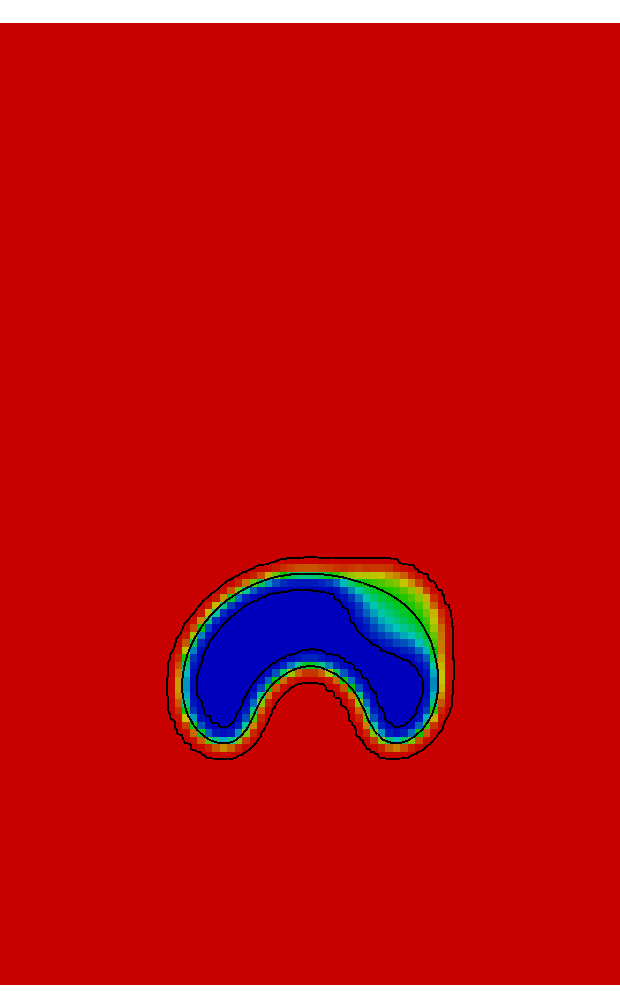}
	\end{minipage}%
	\begin{minipage}{.16\textwidth}
		\centering\includegraphics[width=0.95\textwidth,height=1.5\textwidth,angle=0]{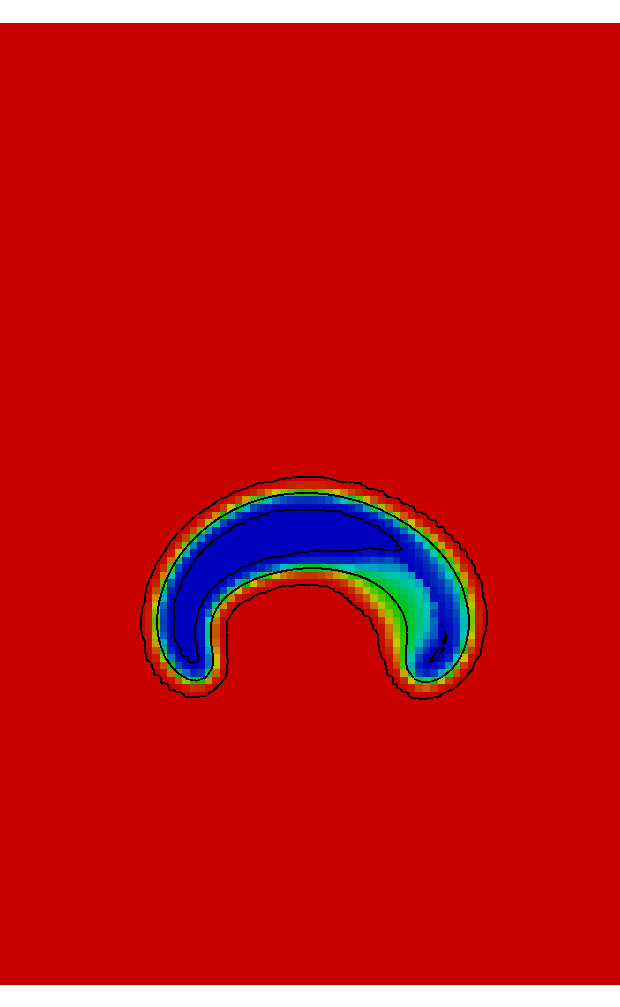}
	\end{minipage}%
	\begin{minipage}{.16\textwidth}
		\centering\includegraphics[width=0.95\textwidth,height=1.5\textwidth,angle=0]{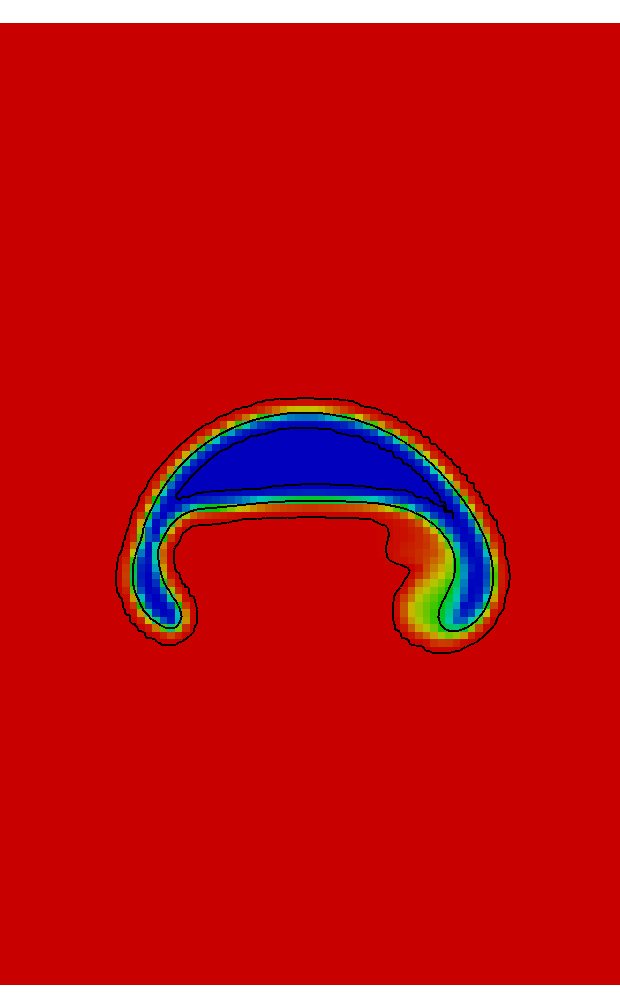}
	\end{minipage}%
	\begin{minipage}{.16\textwidth}
		\centering\includegraphics[width=0.95\textwidth,height=1.5\textwidth,angle=0]{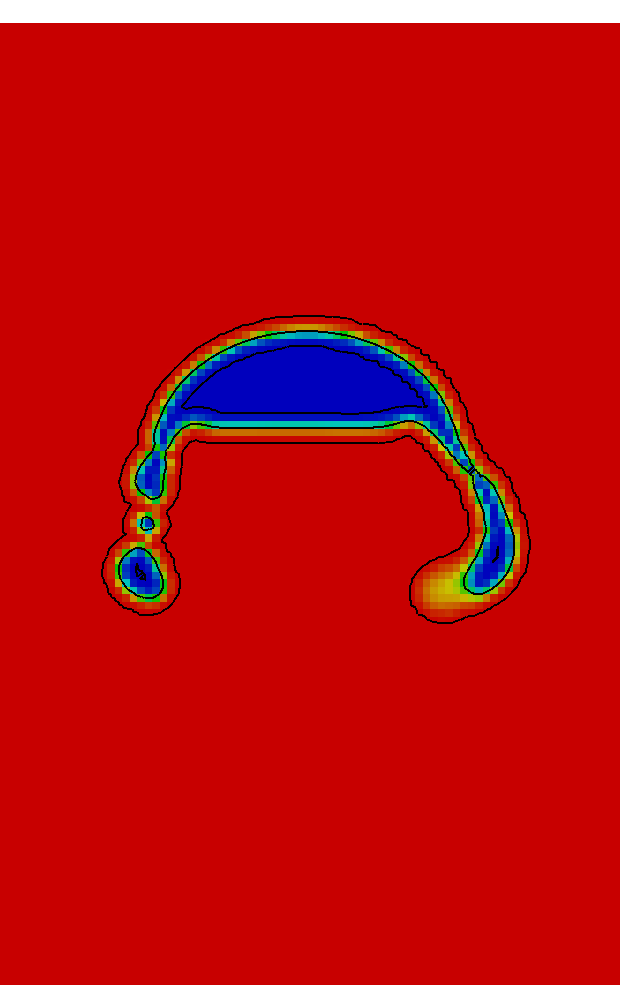}
	\end{minipage}%
	\begin{minipage}{.16\textwidth}
		\centering\includegraphics[width=0.95\textwidth,height=1.5\textwidth,angle=0]{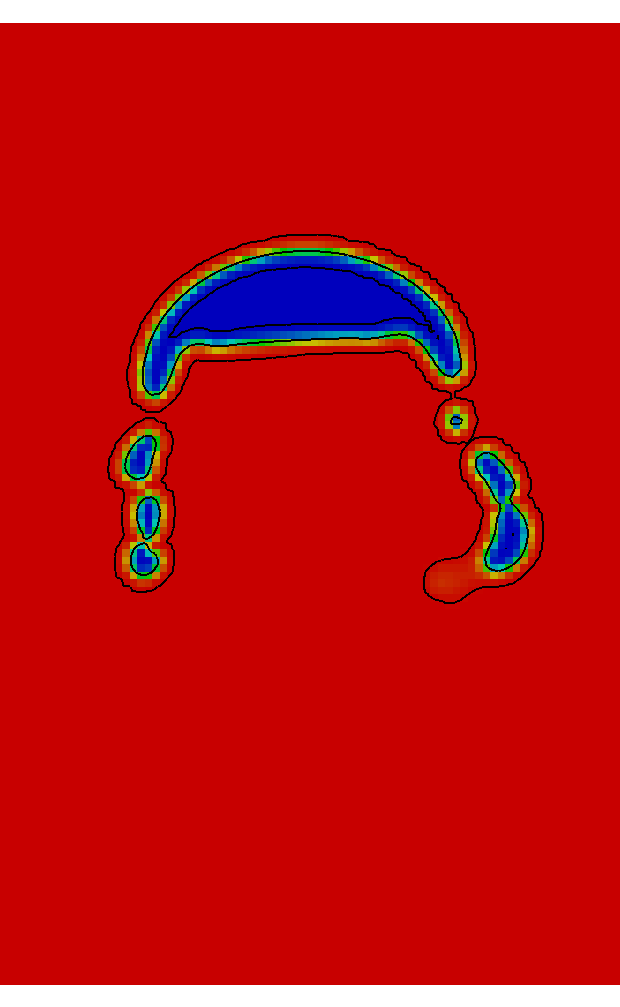}
	\end{minipage}%
	\begin{minipage}{.15\textwidth}
		\includegraphics[width=0.2\textwidth,height=1.25\textwidth,angle=0]{./pics/alpha_bar.png}
	\end{minipage}
	\caption{Level-sets $\ls 0.01,0.5,0.99 \rs$ and color maps 
		of the cumulative distribution functions: $0\!<\!\alpp\!<\!1$ 
		(white contours, top row) and  
		$0\!<\!\alppd\!<\!1$ 
		(black contours, bottom row).
		Position of both functions is 
		affected by the characteristic length
		scale field $\ephxt\,[m]$ with $\ephm\!=\!11\ephb\,[m]$, 
		see \Fig{fig3} (bottom row).
		The snapshots from the left
		to the right are taken every $\Delta t \!=\! 0.05\,[s]$.}
	\bp
	\put(4,271){{{\colw{\small{t=0.05\,[s]}}}}}
	\put(84,271){{{\colw{\small{t=0.10\,[s]}}}}}
	\put(164,271){{{\colw{\small{t=0.15\,[s]}}}}}
	\put(244,271){{{\colw{\small{t=0.20\,[s]}}}}}
	\put(324,271){{{\colw{\small{t=0.25\,[s]}}}}}
	\put(404,271){{{\colw{\small{t=0.30\,[s]}}}}}
	\ep
	\label{fig2}
\end{figure*}
\begin{figure*}[t]
	\begin{minipage}{.16\textwidth}
		\centering\includegraphics[width=0.95\textwidth,height=1.5\textwidth,angle=0]{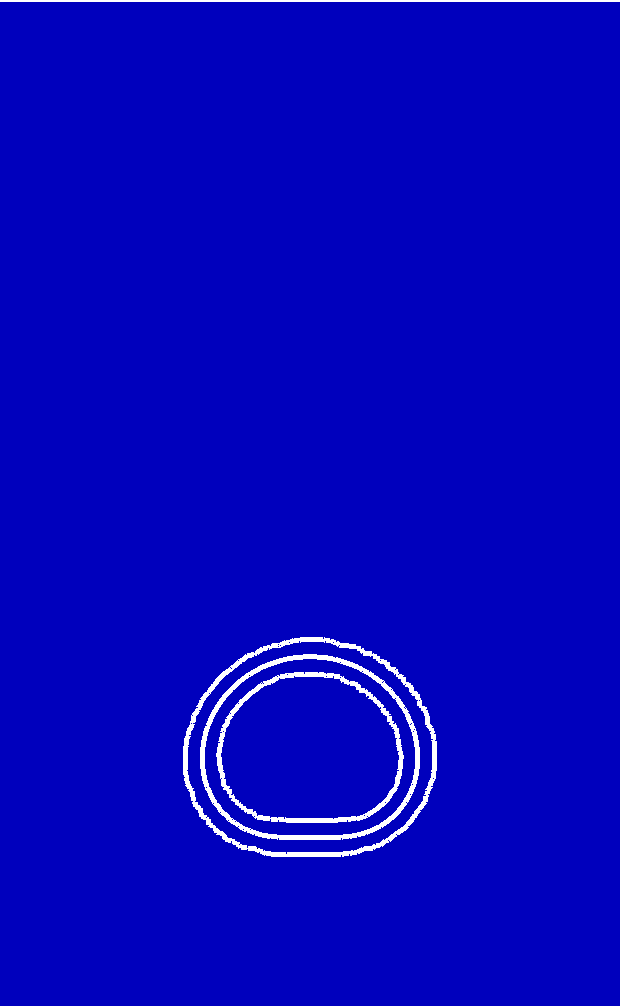}
	\end{minipage}%
	\begin{minipage}{.16\textwidth}
		\centering\includegraphics[width=0.95\textwidth,height=1.5\textwidth,angle=0]{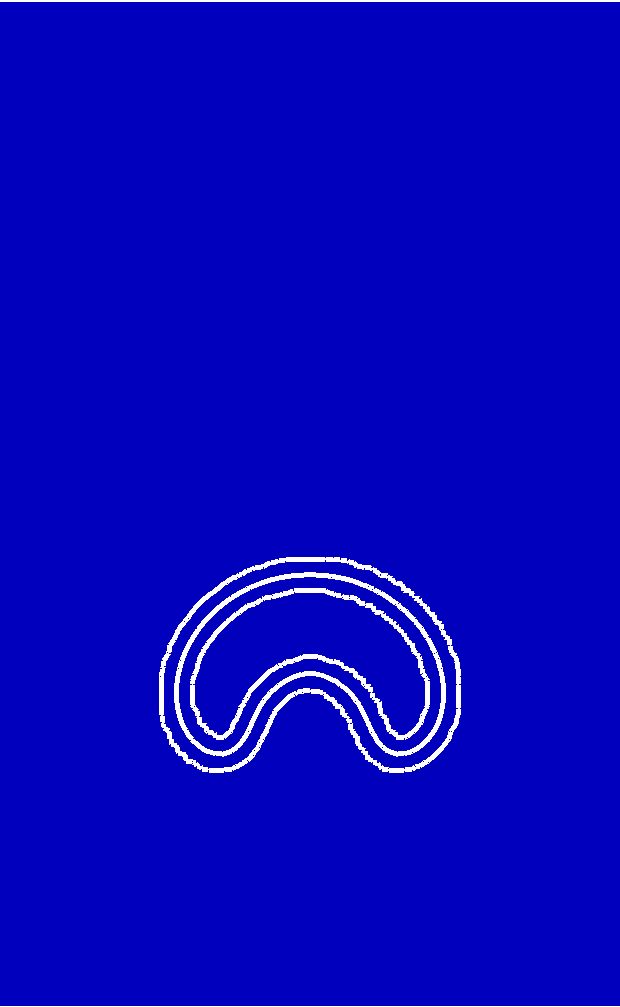}
	\end{minipage}%
	\begin{minipage}{.16\textwidth}
		\centering\includegraphics[width=0.95\textwidth,height=1.5\textwidth,angle=0]{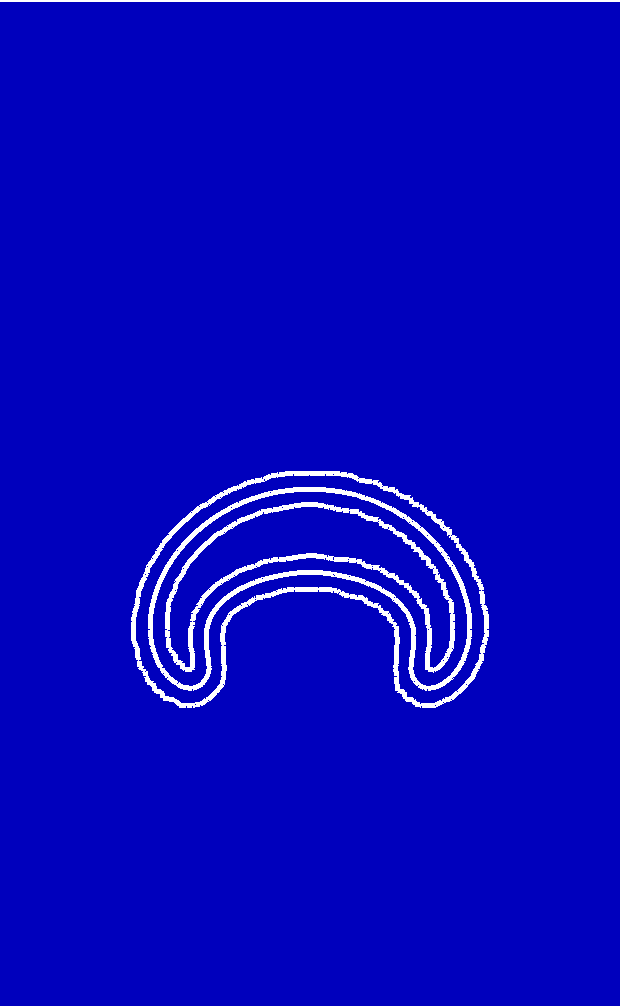}
	\end{minipage}%
	\begin{minipage}{.16\textwidth}
		\centering\includegraphics[width=0.95\textwidth,height=1.5\textwidth,angle=0]{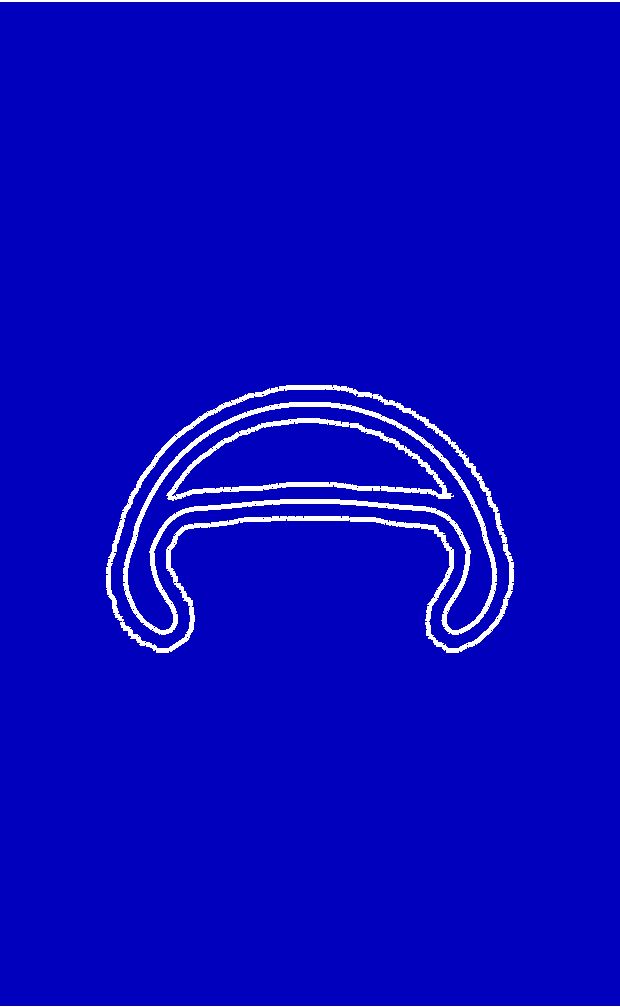}
	\end{minipage}%
	\begin{minipage}{.16\textwidth}
		\centering\includegraphics[width=0.95\textwidth,height=1.5\textwidth,angle=0]{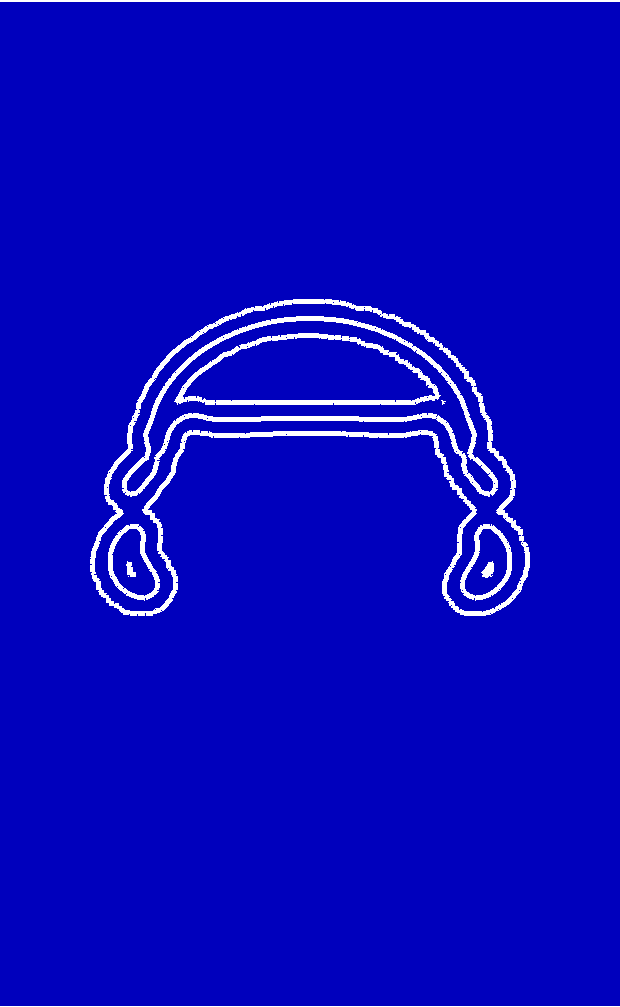}
	\end{minipage}%
	\begin{minipage}{.16\textwidth}
		\centering\includegraphics[width=0.95\textwidth,height=1.5\textwidth,angle=0]{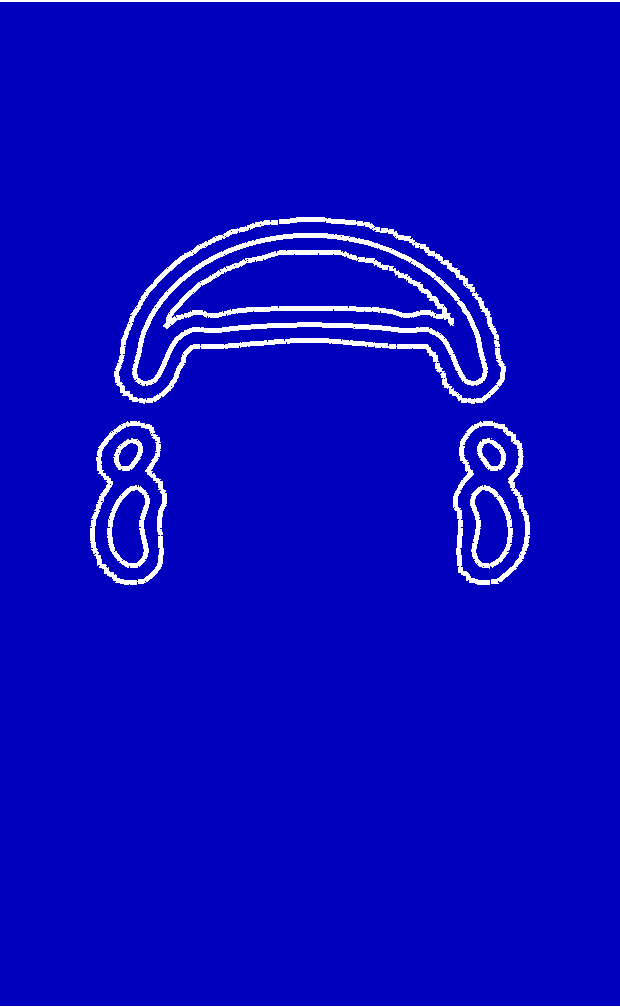}
	\end{minipage}%
	\begin{minipage}{.15\textwidth}
		\includegraphics[width=0.2\textwidth,height=1.25\textwidth,angle=0]{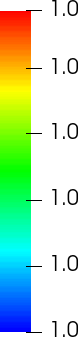}
	\end{minipage}
	\begin{minipage}{.16\textwidth}
		\centering\includegraphics[width=0.95\textwidth,height=1.5\textwidth,angle=0]{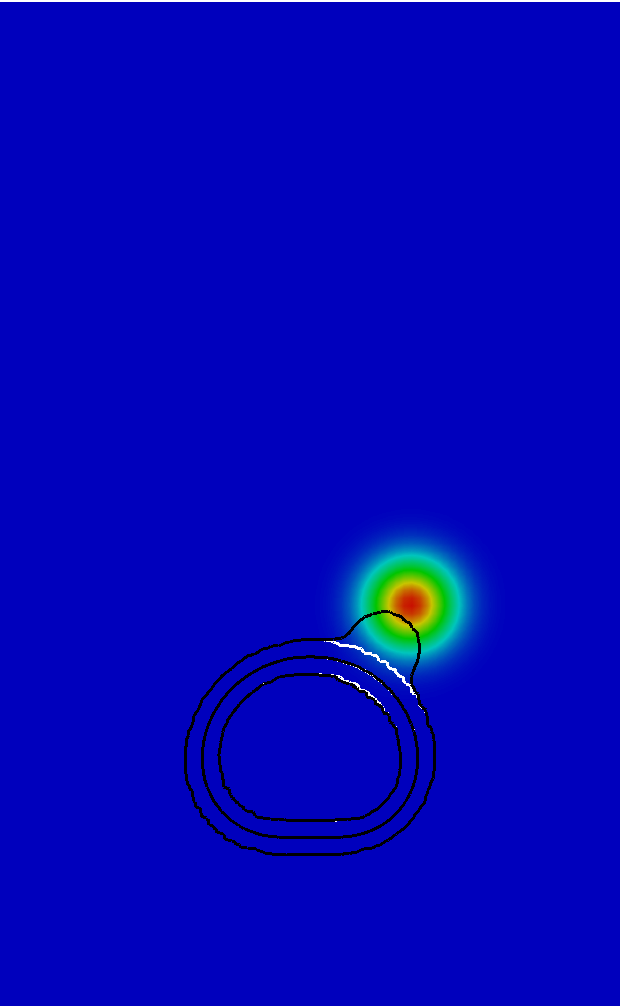}
	\end{minipage}%
	\begin{minipage}{.16\textwidth}
		\centering\includegraphics[width=0.95\textwidth,height=1.5\textwidth,angle=0]{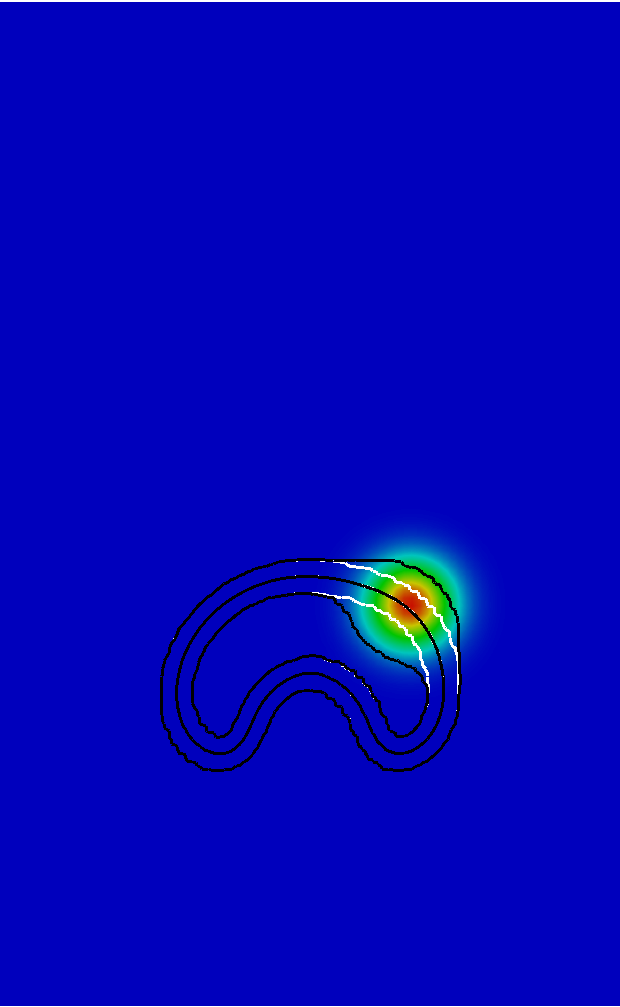}
	\end{minipage}%
	\begin{minipage}{.16\textwidth}
		\centering\includegraphics[width=0.95\textwidth,height=1.5\textwidth,angle=0]{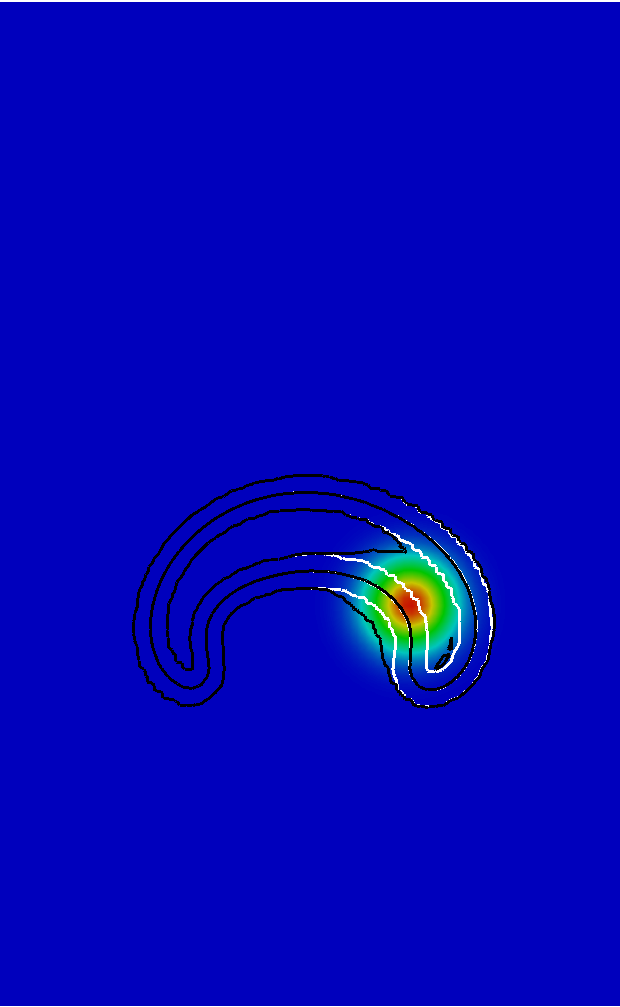}
	\end{minipage}%
	\begin{minipage}{.16\textwidth}
		\centering\includegraphics[width=0.95\textwidth,height=1.5\textwidth,angle=0]{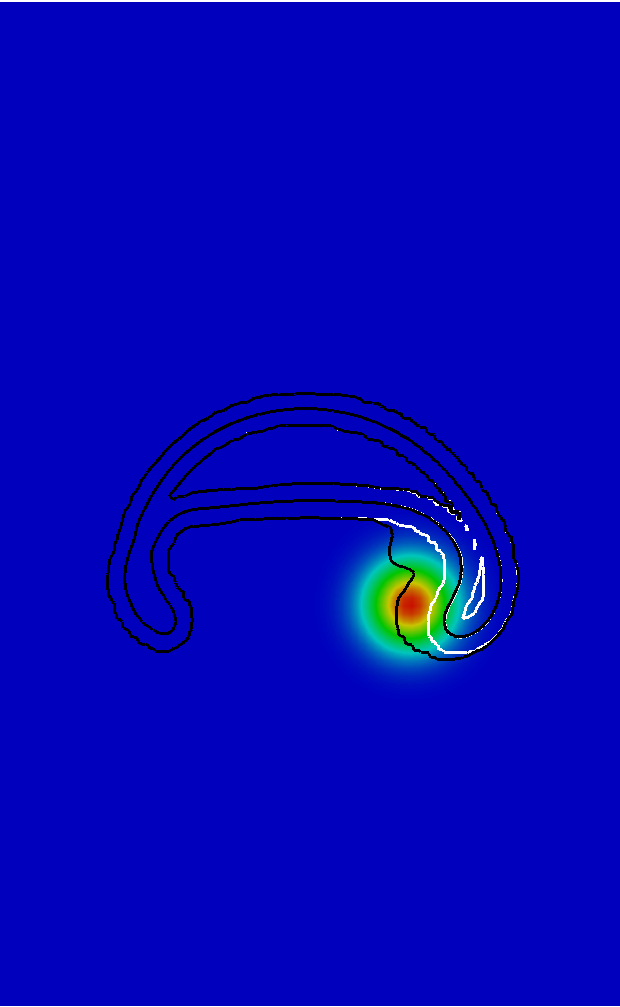}
	\end{minipage}%
	\begin{minipage}{.16\textwidth}
		\centering\includegraphics[width=0.95\textwidth,height=1.5\textwidth,angle=0]{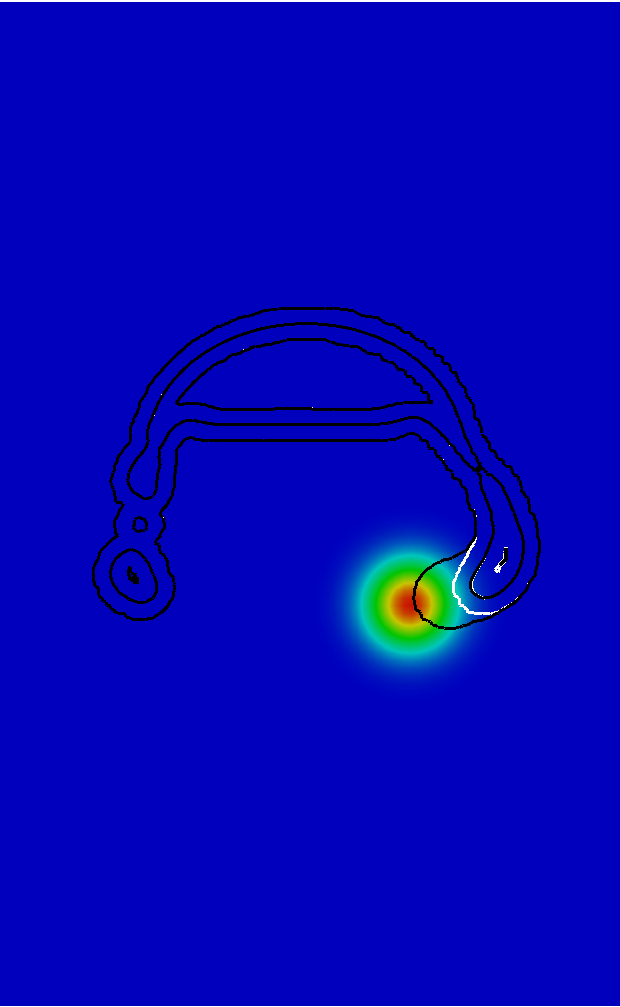}
	\end{minipage}%
	\begin{minipage}{.16\textwidth}
		\centering\includegraphics[width=0.95\textwidth,height=1.5\textwidth,angle=0]{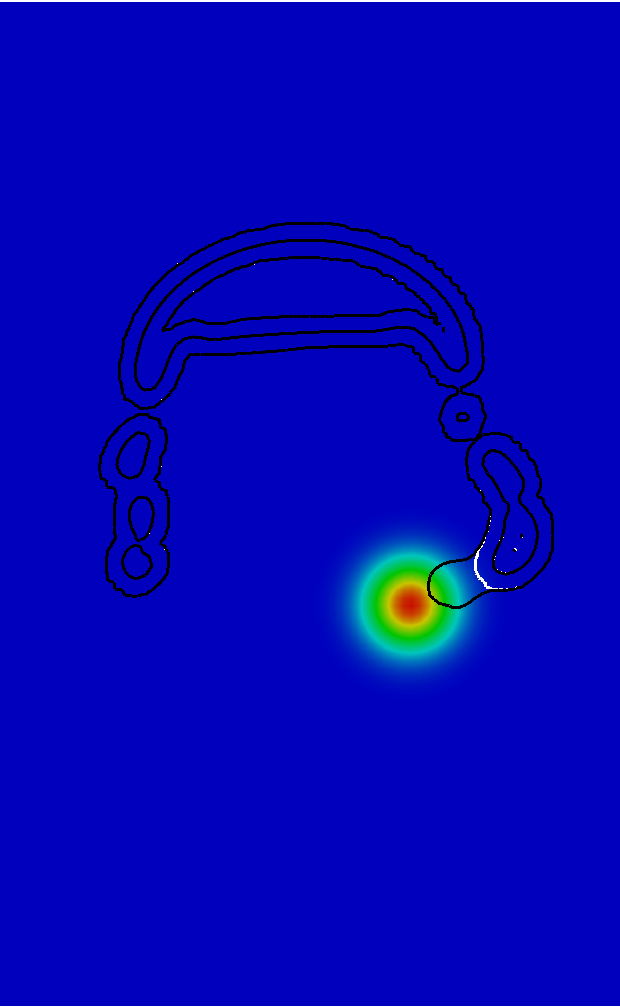}
	\end{minipage}%
	\begin{minipage}{.15\textwidth}
		\includegraphics[width=0.2\textwidth,height=1.25\textwidth,angle=0]{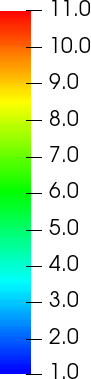}
	\end{minipage}
	\caption{Level-sets $\ls 0.01,0.5,0.99 \rs$ of the probability $\alpp$ with $\ephb\!=\!const.$
		(white contours) and function $\alppd$ 
		approximating the exact solution 
		of \Eq{eq9} (black contour).
		$\alppd$ is
		affected by the characteristic length
		scale field $\ephxt\,[m]$ (colour map).
		$\ephm\!=\!\ephb\!=\!const.$ (top row),
		$\ephm\!=\!11\ephb$ (bottom row).
		The snapshots from left
		to right are taken every $\Delta t \!=\! 0.05\,[s]$.}
	\bp
	\put(4,273){{{\colw{\small{t=0.05\,[s]}}}}}
	\put(84,273){{{\colw{\small{t=0.10\,[s]}}}}}
	\put(164,273){{{\colw{\small{t=0.15\,[s]}}}}}
	\put(244,273){{{\colw{\small{t=0.20\,[s]}}}}}
	\put(324,273){{{\colw{\small{t=0.25\,[s]}}}}}
	\put(404,273){{{\colw{\small{t=0.30\,[s]}}}}}
	\ep
	\label{fig3}
\end{figure*}
%
%
\subsection{Non-equilibrium effects in the two-phase flow}
\label{ssec32}

In the case
when $\ephxt\!=\!const.$,
and the regularized
interface velocity $\bw$
in \Eq{eq8} is prescribed
by  known analytical formula,
the numerical solution
of the transport equations \Eqs{eq8}{eq9}
with constraint given by \Eq{eq4}
are described in 
every detail 
in \cite{twacl15}.
In the present work,
the non-equilibrium
solution of \Eq{eq9},
approximated
by \Eq{eq15} 
is implemented in 
described
in the previous
section
two-phase
flow solver.  
To account for
non-equilibrium
due to variations
in the characteristic
length scale field $\ephxt$,
first, \Eqs{eq8}{eq9} are
solved
with $\ephb\!=\!const.$,
see results presented
in the top rows of
\Figs{fig2}{fig4}
and in the bottom
row of \Fig{fig3}
(white contours). 
This solution provides
$\psia$ and $\alpp$
fields at every
time step.
Further,
$\psia$ is
considered
to define
the local 
coordinate systems
(where normal in
 every point of 
 the interface $\gamma:$
 $\psi \lr \alpha\!=\!1/2 \rr \!=\! 0$ 
 is given by $\bng$).
These local
systems are
used to compute
the integral
\Eq{eq18} in \Eq{eq15}.
The field created
by these local
coordinate systems
is visualized
in \Fig{fig4}
as the level-sets
of $\psia$ 
function.

In \cite{twacl21},
 physical
interpretation of $\psia$
field was
proposed.
It is interpreted
as the space where 
the averaged oscillations
of the mesoscopic, sharp 
interface $\Gamma$
take place.
The 
oscillations 
of $\Gamma$ (in the present work) defined
at the molecular level,
are governed by
thermal fluctuations
and for this reason
can not be
modeled
directly in
the continuum
mechanical
description.
\Eq{eq15}
mimics their
variation in
the averaged
sense, 
see \cite{twacl21}
for full discussion.

Let us consider
the last
outer iteration of
the SIMPLE algorithm, 
after the  
single time 
step $\Delta t\,[s]$
were
the momentum conservation
and Poisson
equations are
iterated to 
enforce stronger
coupling between
the pressure
and velocity
fields.
At the last outer iteration,
$N_\tau\!=\!4$ re-initialization
steps $\Delta \tau \!=\! D/C^2\,[s]$
is carried out
advancing \Eq{eq9}
in time $\tau$.
As discussed
in \Sec{ssec22}, 
this is equivalent
to the minimization 
of the modified
Ginzburg-Landau 
free energy
functional.

Before the
solution of \Eq{eq9}
in time $\tau$, 
the integral
\be
\begin{split}
&\Ip \!=\! \int^0_1 \frac{dt'}{\ephtpxt}  \! \approx \! \\
 &\qquad\qquad \frac{1}{6} \ls \frac{1}{ \eph \texttt{(inp)} } 
+ \frac{4}{ \eph \texttt{(inm)} }
+ \frac{1}{ \eph \texttt{(int)}} \rs.
\label{eq18}
\end{split}
\ee
is computed
in each control 
volume \texttt{(inp)} where the
signed distance
function $\psia$
is reconstructed.
Other
Indexes in \Eq{eq18}
denote, respectively, 
control volumes where:
the interface $\alpha \lr \psi \!=\!0 \rr \!=\! 1/2$ is located $\texttt{(int)}$
and in the distance in between $\texttt{(inm)}$.

It was checked
in \cite{twacl21},
approximation
of the  integral (\ref{eq18})
using the third-order
accurate
Simpson rule
is sufficient
to obtain
$\Ip$ field.
Therein, more detailed
description of the integration
scheme used to evaluate $\Ip$
field is given, too.
Next, obtained
in each control
volume values of $\Ip$
are used to
compute $\alppd$
field given
by \Eq{eq15}
and than material
properties are
determined setting
$\alpp\!=\!\alppd$
in 
\Eqs{eq16}{eq17}.
Afterwards,
the mass
and diffusive fluxes
are calculated
at the faces of all
control volumes.
%
%
Finally,
the segregated
Navier-Stokes equation
solver proceeds
to the solution
on the next time
step $t\!+\!\Delta t$ 
solving euqations
for velocity components
and repeating
described 
procedure.

\section{Numerical experiment}
\label{Sec4}

In this section,
simple two-phase
flow experiment
is carried out
to demonstrate 
the first application 
of the numerical method \cite{twacl21}
used for the approximate solution
of Eqs.~(\ref{eq4}, \ref{eq8}, \ref{eq9})
with \Eq{eq15} which
can account for
the non-equilibrium
effects due to
$\ephxt \! \ne  \! const.$
In particular,
during discussion 
of the numerical results obtained
in this study,
the differences
 between 
statistical and deterministic
description of the two-phase 
flow separated by the
intermittency region
in the equilibrium
or/and non-equilibrium
states
are emphasized.

\subsection{Simulation set-up}
\label{sec41}

In what follows, 
the two-phase
flow of
the rising
due to positive buoyancy
gas bubble is
studied.
For this 
demonstration,
density and viscosity
contrasts
are selected to be
$\rho_1/\rho_2\!=\!6$ and
$\mu_1/\mu_2\!=\!20$,
respectively.
The gravitational
acceleration is chosen
to have its standard magnitude 
$\bg\!=[0,9.8]\,[m/s^2]$.
The two-phase
flow problem
is solved
in the domain
$\Omega = <0,0.1>\!\times\!<0,0.1>\,[m^2]$,
 discretized
using
uniform grid with
$2^7\times2^7$ 
control volumes.
Time
step size
is set to
$\Delta t \!=\! 2.5\cdot10^{-4}\,[s]$
what guarantees
the maximum Courant
number $Cu_{max} \!<\! 0.25\,[-]$
during the whole
simulation lasting
$T\!=\![0,0.5]\,[s]$. 

To couple the pressure
and velocity fields,
$N_{out} \!=\! 4$
outer iterations
per time step 
$\Delta t$ are used.
The same time step 
$\Delta t\,[s]$ is used
to advance
\Eq{eq8} in time $t\,[s]$.
Re-initialization equation (\ref{eq8})
is solved in time $\tau\,[s]$ 
using the time
step size 
$\Delta \tau \!=\! 0.5 \!\cdot\! D/C^2 \!=\! 0.5 \!\cdot\! \ephb\,[s]$
and  $N_\tau\!=\!4$ iterations per $\Delta \tau$.
Model function $\alppd$ given by \Eq{eq15}
is computed after each
re-initialization cycle.
\begin{figure*}[t]
	\begin{minipage}{.16\textwidth}
		\centering\includegraphics[width=0.95\textwidth,height=1.5\textwidth,angle=0]{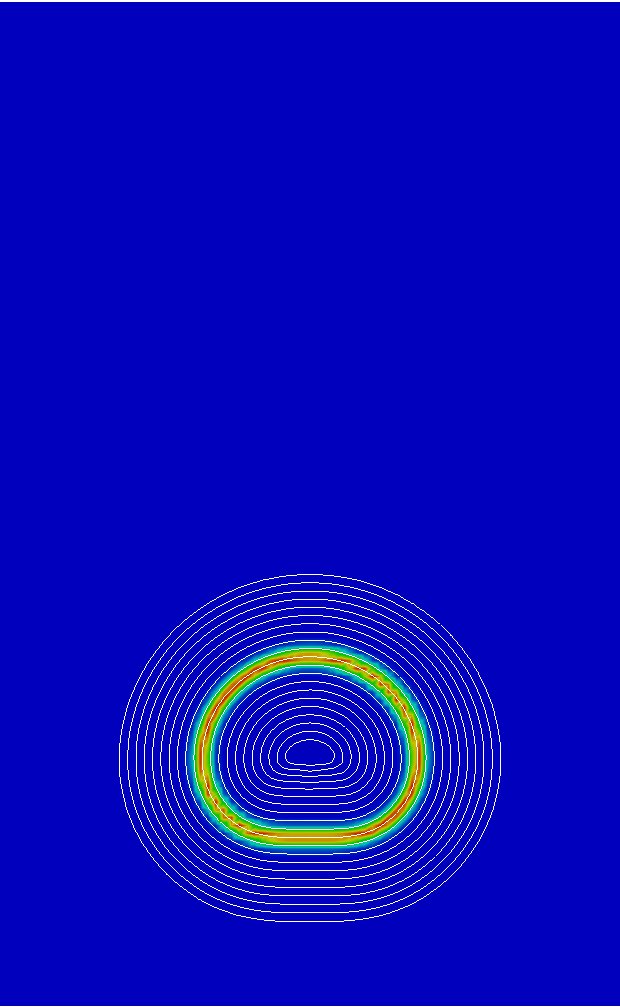}
	\end{minipage}%
	\begin{minipage}{.16\textwidth}
		\centering\includegraphics[width=0.95\textwidth,height=1.5\textwidth,angle=0]{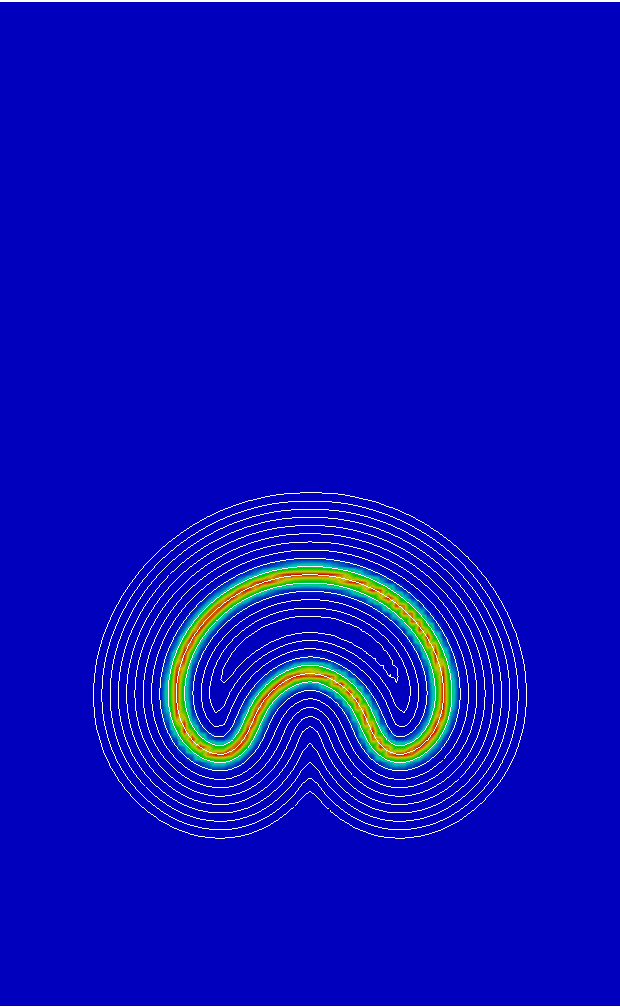}
	\end{minipage}%
	\begin{minipage}{.16\textwidth}
		\centering\includegraphics[width=0.95\textwidth,height=1.5\textwidth,angle=0]{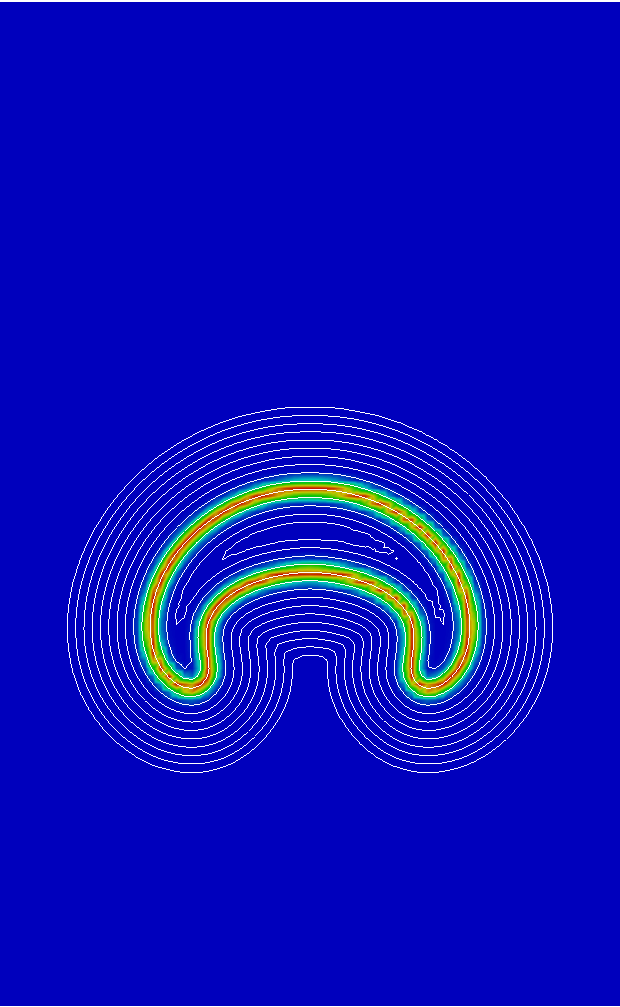}
	\end{minipage}%
	\begin{minipage}{.16\textwidth}
		\centering\includegraphics[width=0.95\textwidth,height=1.5\textwidth,angle=0]{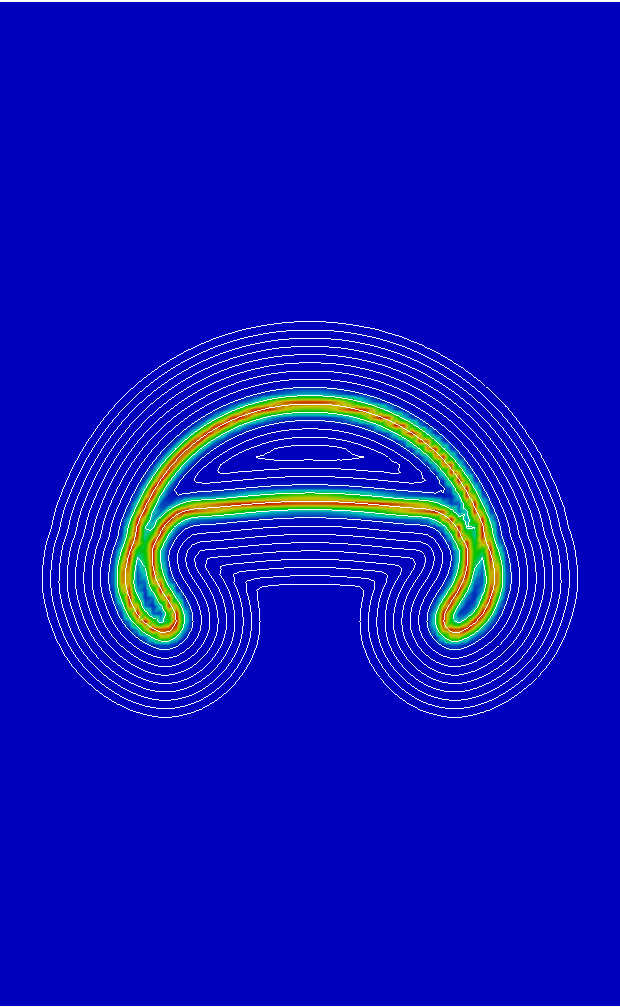}
	\end{minipage}%
	\begin{minipage}{.16\textwidth}
		\centering\includegraphics[width=0.95\textwidth,height=1.5\textwidth,angle=0]{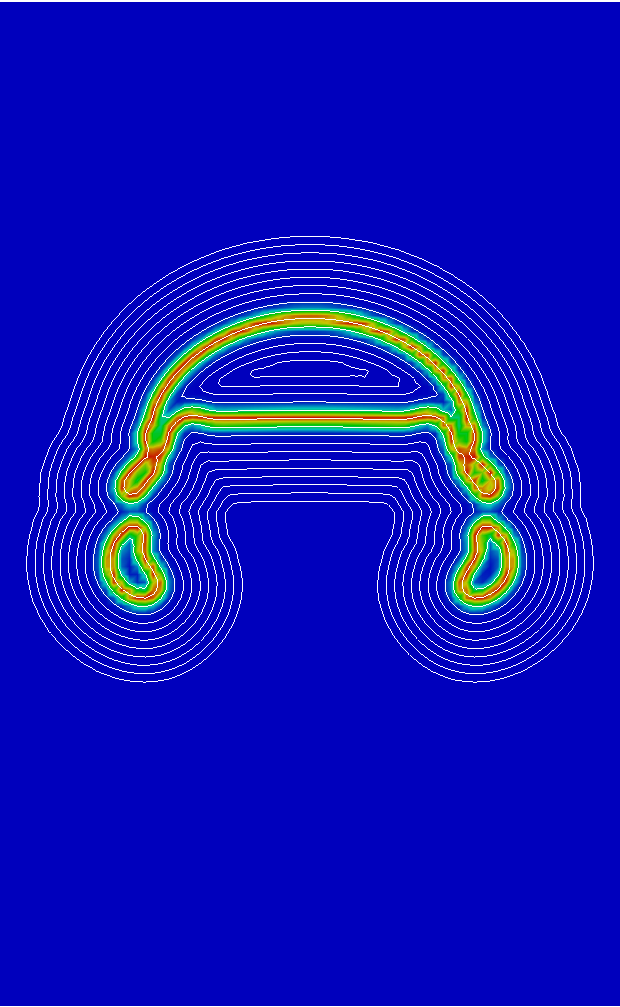}
	\end{minipage}%
	\begin{minipage}{.16\textwidth}
		\centering\includegraphics[width=0.95\textwidth,height=1.5\textwidth,angle=0]{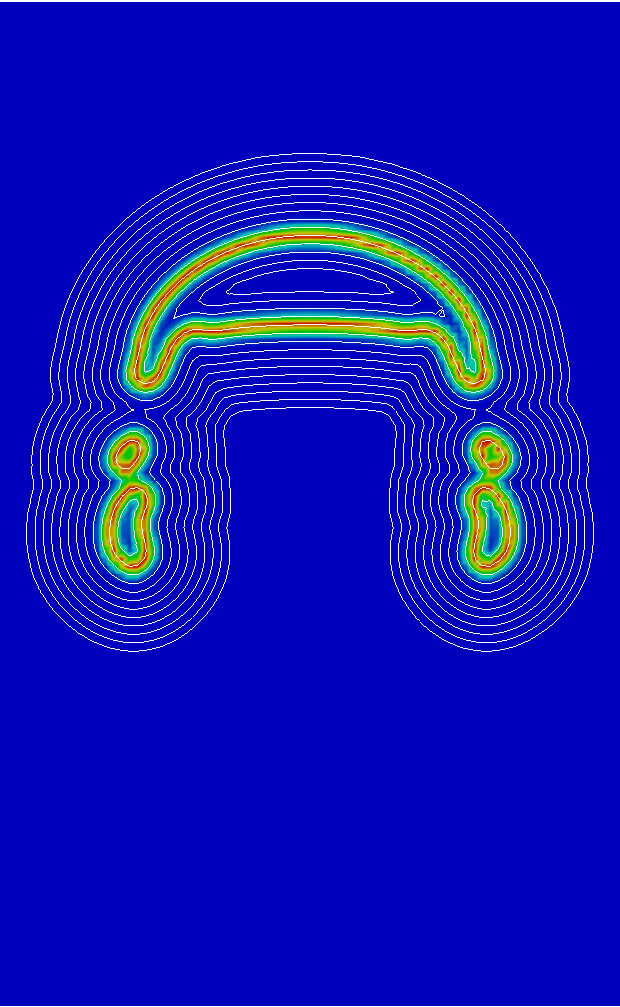}
	\end{minipage}%
    \begin{minipage}{.15\textwidth}
    	\includegraphics[width=0.2\textwidth,height=1.25\textwidth,angle=0]{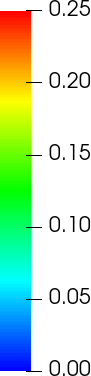}
    \end{minipage}
	\begin{minipage}{.16\textwidth}
		\centering\includegraphics[width=0.95\textwidth,height=1.5\textwidth,angle=0]{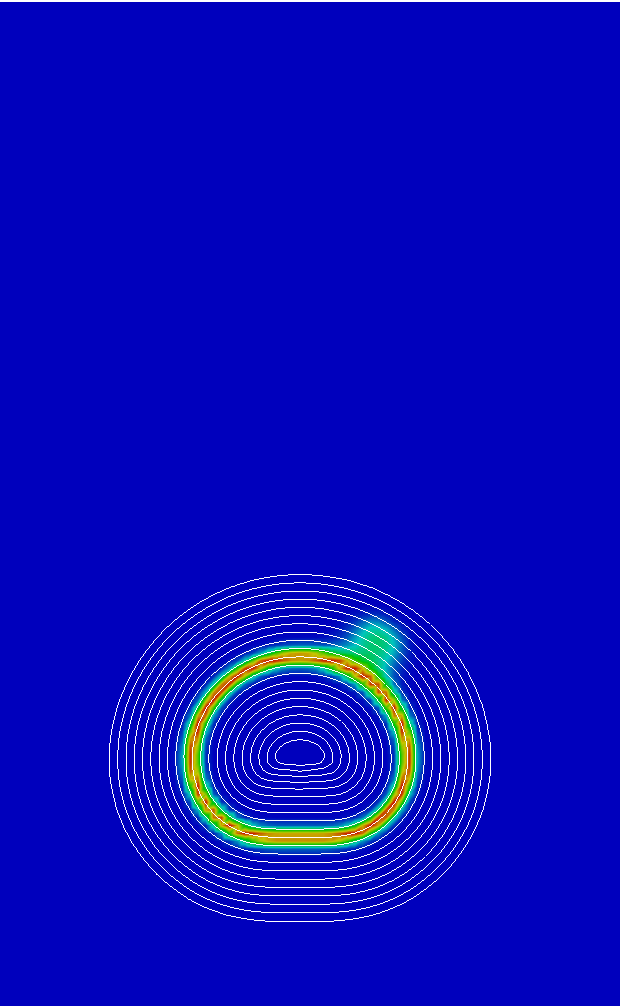}
	\end{minipage}%
	\begin{minipage}{.16\textwidth}
		\centering\includegraphics[width=0.95\textwidth,height=1.5\textwidth,angle=0]{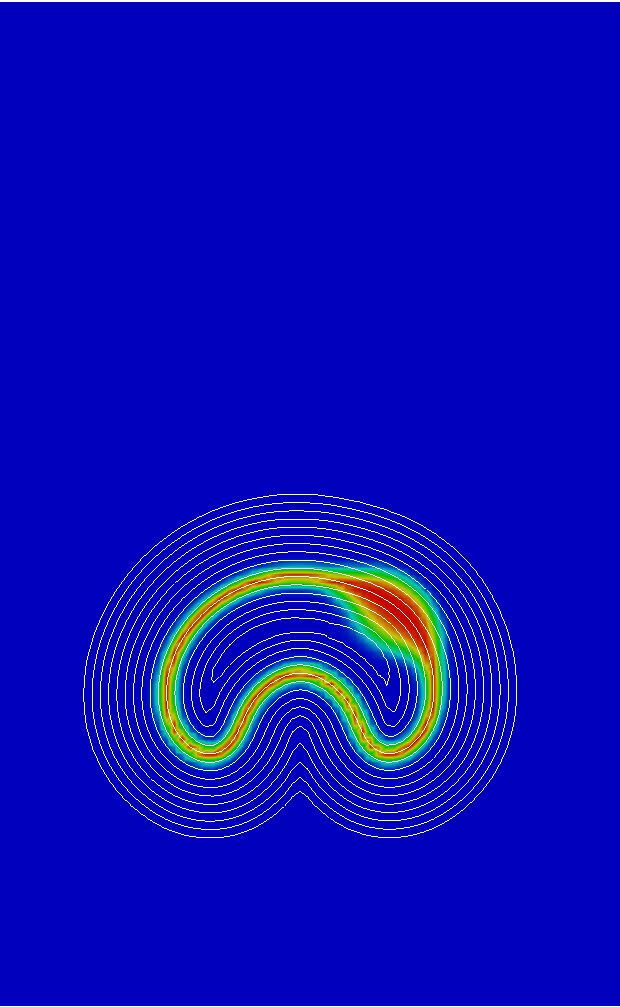}
	\end{minipage}%
	\begin{minipage}{.16\textwidth}
		\centering\includegraphics[width=0.95\textwidth,height=1.5\textwidth,angle=0]{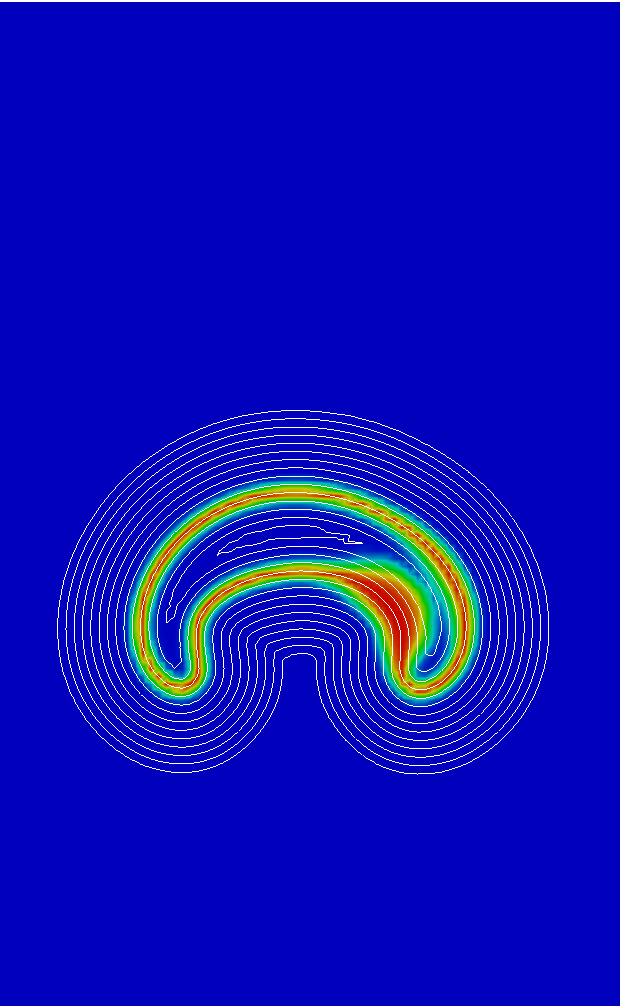}
	\end{minipage}%
	\begin{minipage}{.16\textwidth}
		\centering\includegraphics[width=0.95\textwidth,height=1.5\textwidth,angle=0]{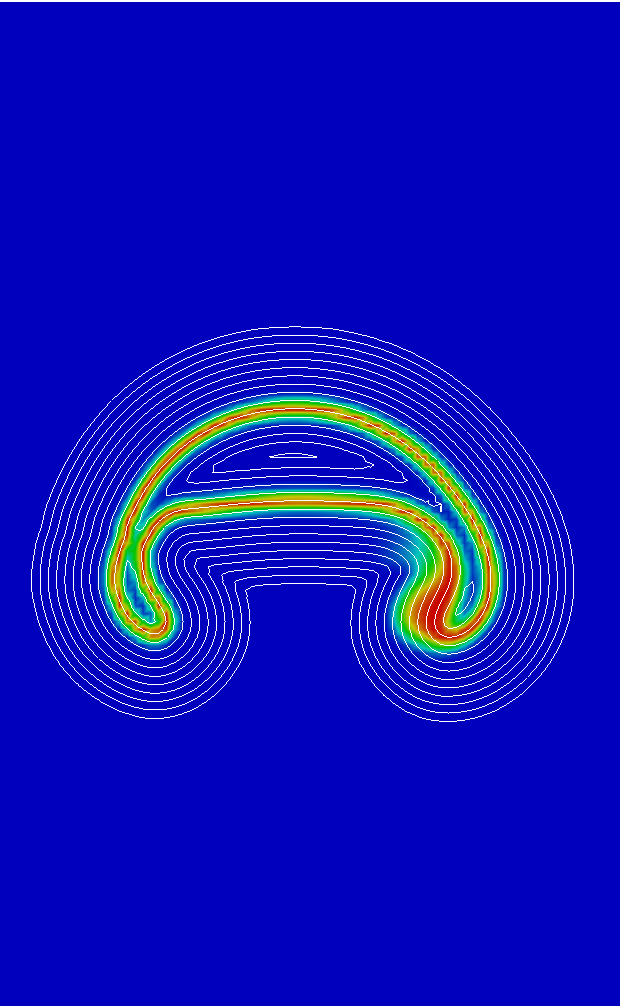}
	\end{minipage}%
	\begin{minipage}{.16\textwidth}
		\centering\includegraphics[width=0.95\textwidth,height=1.5\textwidth,angle=0]{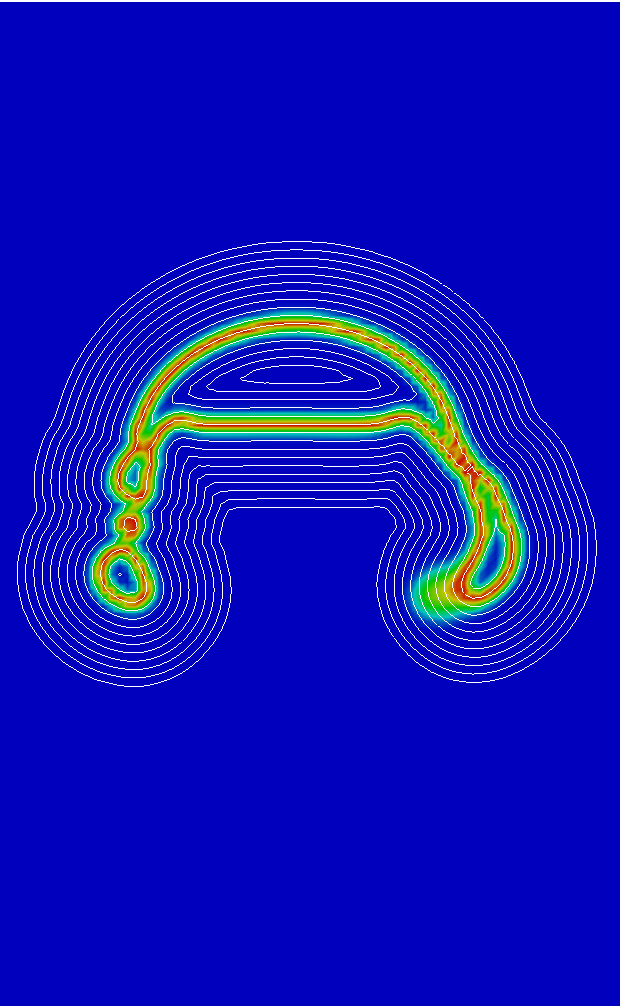}
	\end{minipage}%
	\begin{minipage}{.16\textwidth}
		\centering\includegraphics[width=0.95\textwidth,height=1.5\textwidth,angle=0]{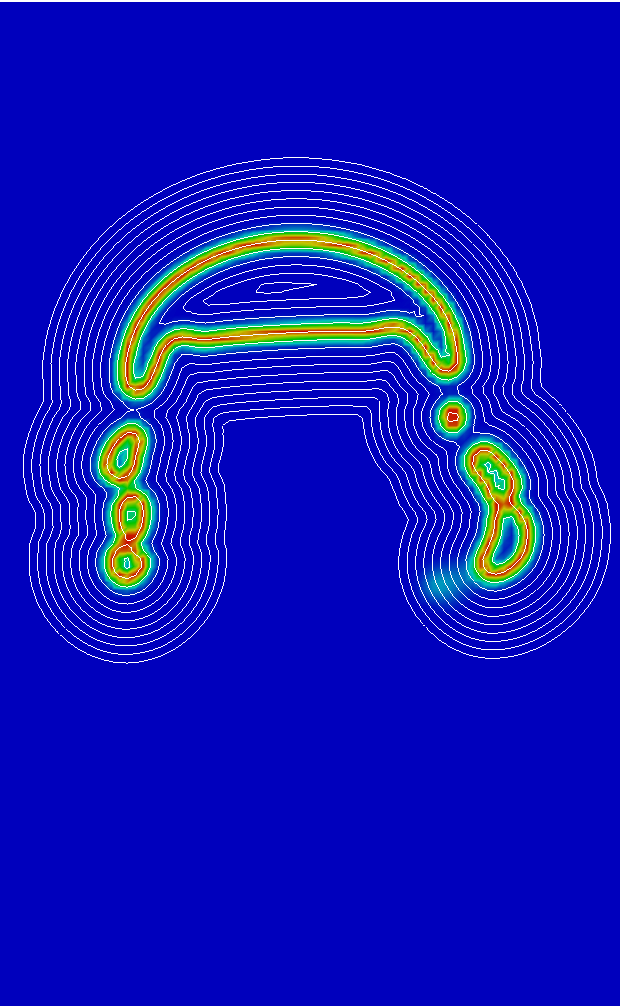}
	\end{minipage}%
	\begin{minipage}{.15\textwidth}
		\includegraphics[width=0.2\textwidth,height=1.25\textwidth,angle=0]{./pics/ndel_bar.png}
	\end{minipage}
	\caption{Level-sets $\lc - 3 \!\cdot\! n \!\cdot\! \ephb, 0, 3 \!\cdot\! n \! \cdot \! \ephb  \rc$ where $n\!=\!1,\ldots,8$ 
		     of the signed distance function field $\psia$.
		     The color map 
			 depicts joint probability $\alppd \lr 1\!-\!\alppd \rr$
		     of finding the phase $\alppd$ used to define the macroscopic interface $\gamma$ position.
			 $\ephm\!=\!\ephb\!=\!const.$ (top row),
			 $\ephm\!=\!11\ephb$ (bottom row).
			 The snapshots from the left
			 to the right are taken every $\Delta t \!=\! 0.05\,[s]$.
		 	}
   	\bp
   	\put(4,273){{{\colw{\small{t=0.05\,[s]}}}}}
   	\put(84,273){{{\colw{\small{t=0.10\,[s]}}}}}
   	\put(164,273){{{\colw{\small{t=0.15\,[s]}}}}}
   	\put(244,273){{{\colw{\small{t=0.20\,[s]}}}}}
   	\put(324,273){{{\colw{\small{t=0.25\,[s]}}}}}
   	\put(404,273){{{\colw{\small{t=0.30\,[s]}}}}}
   	\ep
	\label{fig4}
\end{figure*}

In \Fig{fig1}, 
sketch of the
computational domain
with the initial position  
of the gas bubble $[0.05,0.02]\,[m]$
having diameter 
$D\!=\!0.02\,[m]$
is presented.
The same figure depicts
constant in time position $\bx_r\!=\![0.06,0.04]\,[m]$
of the maximal value
of the characteristic length scale
field $\ephxt$ (colored drop),
see also the color map in \Fig{fig3}. 

The spatial variation
of the characteristic
length scale field
$\ephxt$ is predefined
to be bell shaped
with the analytical formula
\be
\ephxt \!=\! \ephb \lc 1 + A \!\cdot\! \exp{\ls - \lr \frac{r}{20\ephb} \rr^2 \rs} \rc
\label{eq19}
\ee
where $r\!=\!\ls \lr x-x_r \rr^2+\lr y-y_r \rr^2 \rs^{1/2}$,
$\bx_r$ denotes center 
of $\ephxt$,
and the base width of the intermittency
region is  $\ephb=\sqrt{2}\Delta x/4$.

To analyze influence
of $\ephxt$ magnitude
on the obtained results,
in particular 
on the mass
conservation computed
using  $\alpp$ or $\alppd$
functions
several values 
of $A\!=\!\lc 2.5,5.0,7.5,10 \rc$
are used
affecting 
the maximal
elevation of 
$\ephxt$ field.
These $A$ values, result
in  $\ephm \!=\! max \ls \ephxt \rs \!=\!\lc 3.5,6,8.5,11 \rc$
elevations, see 
results in \Fig{fig5}.

It is noticed,
\Figs{fig2}{fig4} 
show only the results
of numerical simulation
in the two cases 
$\ephm/\ephb \!=\! \lc 1,\,11 \rc$.
When $\ephm/\ephb\!=\!1$
than $\ephxt=const.$;
 in the
case $\ephm/\ephb\!=\!11$,   $\ephxt$
is defined by \Eq{eq19}
with $A=10$.

\section{Discusion of the results}
\label{sec5}

\subsection{Conservation of mass}
\label{ssec51}

At first the mass conservation
of the present numerical method
is discussed, see \Fig{fig5}.
As in the present
model we are using
the functions $\alpp$ 
(in the equilibrium case when $\ephxt\!=\!\ephb\!=\!const.$)
and $\alppd$ (to account for variations in $\ephxt$ field),
they both can be used
to compute conservation
of mass or area in
incompressible
two-phase flow.
In the statistical interpretation, 
the mass conservation
is equivalent of the probability 
(of finding phase chosen
to identify the macroscopic 
interface $\gamma$) conservation.

In \Fig{fig5} history of 
the mass variation 
in the present test case
is presented.
This diagram is obtained using
integration of $\alpp$, $\alppd$
functions in the computational
domain there where $\psia$ 
is reconstructed.
It is noted, the total mass/area
is slightly higher then $S_{ext}=\pi D^2/4$
used to normalize data in \Fig{fig5}.
This is because $\alpp$, $\alppd$
are the cumulative distribution functions
with the infinite support and
they are resolved  that
way as accurate as it is possible
with the double precision computer
arithmetic.
In both cases presented
in \Fig{fig5} the law of mass 
conservation is satisfied
during the whole simulation.
In this test, the
coarse mesh is used
($\sim 26$ grid nodes per buble
diameter $D$ at $t=0\,[s]$), 
the relatively exact
mass conservation is due to  
the conservative form
of the transport \Eq{eq7}.

In the case of 
the mass recording 
obtained
using $\alpp$ 
in \Fig{fig5} (left)
one observes 
how coupling 
through continuity
equation with
the non-equilibrium solution 
affects its conservation.
Counter intuitively, 
the larger $\ephxt$
field magnitude $\ephm$ is,
the lower mass error
is obtained.

In \Fig{fig5} (right),
conservation
of $\alppd$ 
is depicted.
Herein, one notes interaction
with the variable characteristic
length scale field $\ephxt$
causes 
fluctuations of
the mass magnitude 
in $\pm3\%$ range.
However, 
due to coupling
with the conserved
$\alpp-\psia$ functions
the total mass 
is conserved as well.

The time moments
of oscillations peaks
in \Fig{fig5} (right) 
can be compared
with the two-phase
flow field(s)
in \Figs{fig2}{fig4}.
One can observe,
these peaks
occur when the non-symmetrical
modes of $\alppd$ are induced
by $\ephxt$ field.
This can be obseved
at times
$t=0.015,0.02\,[s]$,
e.g. compare results
in \Fig{fig5}
and \Fig{fig2}. 

%
\subsection{Impact of non-equilibrium effects on two-phase flow}
\label{ssec52}
%

Impact of the characteristic
length scale field $\ephxt$ on 
the obtained results is presented
in \Figs{fig3}{fig4}, 
compare results in the top
and bottom rows.
One observes,
in the case $\ephxt=\ephb$
investigated two-phase flow
is symmetric.
In spite they
are there, black
contours of $\alppd$
field can not be seen
in \Fig{fig3} (top row)
because in this 
case $\alppd\!=\!\alpp$.

This is not the case
in the bottom row of
\Fig{fig3}.
Herein, the difference
between $\alpp$ (white contours)
and $\alppd$ (black contours)
functions is clearly
visible.
As in the non-equilibrium case, 
the material properties of
gas and liquid phases
are computed using 
\Eqs{eq15}{eq16} with
$\alpp$ replaced by $\alppd$.
This locally changes
the mass and diffusive
fluxes used to solve
the Navier-Stokes equations
and couple
the pressure
and velocity
fields.
Therefore now,
the two-phase flow
scenario is
affected by $\alppd$ 
variations 
causing
the flow
symmetry
to be broken.
This can be
attributed to
differences in
topological changes
on the left and
right side of 
the gas bubble,
compare results
in the top and bottom 
rows in \Figs{fig3}{fig4}
for example
at times $t=0.25,\,0.3\,[s]$.

In our opinion,
the best illustration
of mechanisms in 
the statistical two-phase 
flow model is given
by the joint probabilities
$\alpp \lr 1-\alpp \rr$ 
or
$\alppd \lr 1-\alppd \rr$ 
depicted in the top
and bottom rows in \Fig{fig4}, 
respectively.
%
%
%
Variations 
of the joint probability
used to identify macroscopic
interface $\gamma$ as $\psi \lr \alpha=1/2 \rr=0$,
illustrate
the probability distribution
of the mesoscopic 
interface $\Gamma$
residence.
In other words,
in regions with higher
joint probability, it is
more likely to find the
mesoscopic interface $\Gamma$
(we recall in the present
work $\Gamma$ is defined on the 
molecular level).

\subsection{Difference between statistical and deterministic two-phase flow models}
\label{ssec53}

Finally,
let us  discuss
the main
difference
between
deterministic 
and statistical \cite{mwaclawczyk11,waclawczyk2015,twacl15,twacl17,twacl21}  two-phase flow
models used in the one-fluid model
framework.  

In the  deterministic description
of the interface,
the physical interpretation
is attributed only to flow
domains where the order parameter or
the phase indicator function
has value zero or one \cite{trygg11,kim12}.
According to models of
Gibbs \cite{gibbs1874} and
van der Waals \cite{waals1979},
these domains are occupied
by the homogeneous gas or liquid
phases \cite{lang15}.
However, 
in 
physical/chemical reality,
gases or liquids will never 
be homogeneous 
(gas or liquid phases containing
100\% of one type molecules
are almost never observed in every
day experience). 
Moreover,
this constraint imposed on
the deterministic sharp interface models
results in the presumption
that
 domains where 
 the numerical 
 solution of the order parameter
 or phase indicator transport 
 equation is between zero and one
 (smeared, not
sharply resolved regions)
are non-physical.
Similarly,
diffusive interface 
in the phase field models
is considered to be thin
transition region 
between the  two-phases
without clear  physical
interpretation as well.
Nevertheless,
exceptionally,
the position
of the interface
in its sharp and/or
diffusive deterministic 
models 
is typically located
between zero and one.
As a consequence,
the only possibility
to increase
the accuracy
of the numerical
predictions
with the sharp or
diffusive
interface model(s) 
is 
addition of 
the new
grid points using
the adaptive
refinement
what impairs
performance
of the solver
(steam is putted
to the wrong, 
non-physical wheel). 

In the statistical model
of  intermittency region,
the cumulative distribution
function describing
the probability
of finding
mesoscopic interface 
$\Gamma$ 
(or  probability
of finding the phase chosen to
identify $\Gamma$ and $\gamma$)
is 
bounded $0\!<\!\alpp\!<\!1$
and  conserved
from its definition,
see mathematical
form of \Eq{eq7}.
In this interpretation,
the expected position of the
interface $\Gamma$ defined as
$\alpha \lr \psi\!=\!0 \rr \!=\! 1/2$
explains where the macroscopic 
interface $\gamma$ is 
located (also
in the deterministic
interface models).
Moreover, values
of $\alpp$ have
the physical interpretation
over the entire
range of $\alpp$
field magnitude.
Hence,
regions
where
the two-phase flow 
is not resolved
due to lack of
the spatial
resolution are
promoted to
physical 
interpretation
as well.
%
%
%
This feature of 
the statistical model
of the intermittency region
is general and applies
equally to the equilibrium
and non-equilibrium flow
cases.
In order not 
to be too vague, 
let us analyze 
the results
obtained 
in the
present numerical
experiment
in the context
of above
explanations.
%
%
%

It is recalled, 
$\alpp$ and $\alppd$
are probabilities of finding
the liquid phase,  
$1\!-\!\alpp$ and $1\!-\!\alppd$
are probabilities of finding
the gas phase.
We note, at $t=0.3\,[s]$ (see \Figs{fig2}{fig4}) some
features of the rising gas bubble are
not well resolved.
Contours of $\alpp$ (white)
and $\alppd$ (black) capture only values
$\lc 0.5,0.99 \rc$, see \Figs{fig2}{fig3}. 
Unlike in the deterministic
interface models,
the present numerical solution,
although not fully resolved,
still carries
a physical 
information.
Namely it
means, that
finding of 
the  gas phase 
in such regions
is not certain
as
 $1\!-\!\alpp \!<\! 0.99$
and this is why 
$\alpp\!=\!0.01$ or  $\alppd\!=\!0.01$
contours can not be seen. 
%
\section{Conclusions}
\label{Conclusions}

In the present paper, 
the statistical model of
the intermittency region
between two phases \cite{twacl17,twacl21}
was coupled with 
the incompressible 
Navier-Stokes equation
solver 
to study 
differences between
the equilibrium and 
non-equilibrium solution
of \Eq{eq7} in 
the case of 
rising gas bubble.
Results
of the present
numerical simulations
show,
inclusion of
recently 
proposed 
approximate
solution to \Eq{eq7} 
permits to account
for non-equilibrium effects
in the
two-phase flow model
rooted in the one-fluid
framework. 

Moreover, we
have discussed 
differences
between deterministic
sharp and diffusive
interface models
and statistical 
interface model.
It is concluded,
the main difference
between 
these two descriptions
of two-phase flow
is that the results of
the statistical
two-phase flow 
model can be interpreted
over the entire range
of the cumulative
distribution function
$\alpp$ values
also when 
two-phase 
flow is
not fully
resolved.


\bibliographystyle{spmpsci}
\bibliography{mybibfile.bib}

\end{document}